\documentclass[acmsmall,screen,urlbreakonhyphens,nonacm]{acmart}

\renewcommand\footnotetextcopyrightpermission[1]{}
\setcopyright{none}

\usepackage[utf8]{inputenc}

\usepackage[capitalize]{cleveref}
\usepackage{amsthm}
\usepackage{mathtools}
\usepackage{physics}
\usepackage{xcolor}
\usepackage{graphicx}
\usepackage{bm}
\usepackage{fontawesome}

\usepackage{relsize}
\usepackage{xspace}
\usepackage{wrapfig}

\newcommand{\nocontentsline}[3]{}
\newcommand{\tocless}[2]{\bgroup\let\addcontentsline=\nocontentsline#1{#2}\egroup}

\newcommand{\linktonotebook}[2]{\NoCaseChange\href{https://github.com/mapequation/infomap-tutorial-notebooks/blob/main/#1.ipynb}{{#2}~\faExternalLink}}

\newcommand{\CC}{C\nolinebreak\hspace{-.05em}\raisebox{.2ex}{\bf +}\nolinebreak\hspace{-.05em}\raisebox{.2ex}{\bf +}}

    \AtBeginDocument{%
  }

\begin{document}

\title{Community Detection with the Map Equation and Infomap: Theory and Applications}

\author{Jelena~Smiljani{\'c}}
\authornote{Both authors contributed equally to this research.}
\email{jelena.smiljanic@umu.se}
\orcid{0000-0003-0124-1909}
\affiliation{%
  \institution{IceLab, Department of Physics, Ume{\aa} University}
  \city{Ume{\aa}}
  \country{Sweden}
  \postcode{SE-901 87}
}
\affiliation{%
  \institution{Scientific Computing Laboratory, Center for the Study of Complex Systems, Institute of Physics Belgrade, University of Belgrade}
  \streetaddress{Pregrevica 118}
  \city{Belgrade}
  \country{Serbia}
  \postcode{11080}
}

\author{Christopher~Bl{\"o}cker}
\authornotemark[1]
\email{christopher.blocker@umu.se}
\orcid{0000-0001-7881-2496}
\affiliation{%
  \institution{IceLab, Department of Physics, Ume{\aa} University}
  \city{Ume{\aa}}
  \country{Sweden}
  \postcode{SE-901 87}
}
\affiliation{%
  \institution{Center for Artificial Intelligence and Data Science, University of W{\"u}rzburg}
  \city{W{\"u}rzburg}
  \country{Germany}
  \postcode{DE-97070}
}
\affiliation{%
  \institution{Data Analytics Group, Department of Informatics, University of Z{\"u}rich}
  \city{Z{\"u}rich}
  \country{Switzerland}
  \postcode{CH-8006}
}

\author{Anton~Holmgren}
\email{anton.holmgren@umu.se}
\orcid{0000-0001-5859-4073}
\affiliation{%
  \institution{IceLab, Department of Physics, Ume{\aa} University}
  \city{Ume{\aa}}
  \country{Sweden}
  \postcode{SE-901 87}
}

\author{Daniel~Edler}
\email{daniel.edler@umu.se}
\orcid{0000-0001-5420-0591}
\affiliation{%
  \institution{IceLab, Department of Physics, Ume{\aa} University}
  \city{Ume{\aa}}
  \country{Sweden}
  \postcode{SE-901 87}
}
\affiliation{%
  \institution{Gothenburg Global Biodiversity Centre, University of Gothenburg}
  \city{Gothenburg}
  \country{Sweden}
}

\author{Magnus~Neuman}
\email{magnus.neuman@umu.se}
\orcid{0000-0002-3599-9374}
\affiliation{%
  \institution{IceLab, Department of Physics, Ume{\aa} University}
  \city{Ume{\aa}}
  \country{Sweden}
  \postcode{SE-901 87}
}

\author{Martin~Rosvall}
\email{martin.rosvall@umu.se}
\orcid{0000-0002-7181-9940}
\affiliation{%
  \institution{IceLab, Department of Physics, Ume{\aa} University}
  \city{Ume{\aa}}
  \country{Sweden}
  \postcode{SE-901 87}
}



\tocless\begin{abstract}
Real-world networks have a complex topology comprising many elements often structured into communities.
Revealing these communities helps researchers uncover the organizational and functional structure of the system that the network represents.
However, detecting community structures in complex networks requires selecting a community detection method among a multitude of alternatives with different network representations, community interpretations, and underlying mechanisms.
This tutorial focuses on a popular community detection method called the map equation and its search algorithm Infomap.
The map equation framework for community detection describes communities by analyzing dynamic processes on the network.
Thanks to its flexibility, the map equation provides extensions that can incorporate various assumptions about network structure and dynamics.
To help decide if the map equation is a suitable community detection method for a given complex system and problem at hand -- and which variant to choose -- we review the map equation's theoretical framework and guide users in applying the map equation to various research problems.
\end{abstract}






\maketitle
\newpage
\addtocontents{toc}{\protect\setcounter{tocdepth}{-1}}
\tableofcontents
\addtocontents{toc}{\protect\setcounter{tocdepth}{2}}
\clearpage
\section{Introduction} \label{sec:introduction}

Networks help researchers analyze complex systems by representing their intricate interactions.
To simplify networks with numerous nodes and links and uncover their essential structure and dynamics, researchers have developed an array of methods to detect communities, also called modules~\cite{fortunato2010physrep, fortunato2016physrep}.
These methods vary in goals, assumptions, and interpretations, ranging from generative models representing network formation processes to flow-based approaches capturing processes on the network~\cite{schaub2017many}.
When choosing a method for a specific problem, users must consider their objectives, the system under study, and the available data~\cite{peel2017ground}.
The multitude of methods and settings can be daunting.
Users often struggle to select the most suitable approach and configuration, leading to suboptimal results.

Two questions stand out when describing network communities: How did the network form?
How does the network's structure affect processes on the network?
While these questions are often intertwined, and community detection methods handle assumptions about the network's formation, structure, and dynamic processes differently, we can discern two major approaches.
Different generative methods address the first question, such as community embedding models~\cite{sun2019vgraph}, community-affiliation graph models~\cite{yang2014overlapping}, and stochastic block models with Bayesian inference techniques~\cite{peixoto2019sbm}.
These methods assume that a latent community structure determines the link distribution.
Finding an optimal partition representing the underlying structure involves applying statistical inference to fit a generative model to the network data.
Addressing the second question requires shifting focus to the processes on the network that uncover how nodes share similar dynamical roles.

In this review, network flows refer to dynamic processes on networks, such as the spread of information, activity, or contagion, which can often be modeled as random walks. This usage differs from the traditional meaning of network flow in computer science and operations research, where it typically refers to capacity-constrained optimization problems such as the max-flow min-cut problem~\cite{ford1956maximal}. Network flows, understood as dynamic processes on networks, are vital for our understanding of system-wide behavior in complex systems. 
They transform networks from combinatorial constructs of nodes and links into integrated representations of complex systems where distant parts can influence each other. 
Specifically, network flows capture the interconnected dynamics of complex systems, such as when element \textsf{A} interacts with element \textsf{B} and element \textsf{B} interacts with element \textsf{C}, elements \textsf{A} and \textsf{C} indirectly affect each other.

Sometimes, these flows are observable, such as passenger flows between airports in air traffic networks.
Other times, we must model the flows.
For example, when we want to comprehend genetic pathways but have access only to the gene network, we can derive flows from the interactions between pairs of genes.
Since network structure constrains processes on it, studying modeled flows can also reveal meaningful structures in networks without explicit flows.

Flow-based modules coarse-grain the dynamics taking place on the network.
Groups of nodes where the network flows stay relatively long provide an intuitive notion of flow-based modules.
In air traffic networks, passengers traveling remain within modules comprising sets of airports that contain numerous travel routes. 
In gene networks with unknown genetic pathways, functional modules trap modeled network flows representing the biological processes.
In networks without explicit network flows, including synthetic benchmark networks with planted community structure of a generative model, flow-based methods have nevertheless proven effective in identifying these topological communities \cite{lancichinetti2009pre,Aldecoa2013,10.1371/journal.pone.0154404}.
Because studying flow-based communities provides essential insights about the systems they represent, reliably identifying them is critical.

While there are several flow-based community detection methods~\cite{pons2005flows,delvenne2010flows,schaub2012plosone,lambiotte2014flows}, we focus on the map equation and its search algorithm Infomap~\cite{rosvall2008pnas,mapequation2022software}.
The map equation captures modular regularities by compressing the description of flows, capitalizing on the minimum description length principle~\cite{rissanen1978modeling}.
Applied to modular compression of network flows, the minimum description length principle states that finding the partition that enables the greatest compression of the network flows is equivalent to identifying the modules that best capture the regularities of those flows.
Given a partition, the map equation calculates the lower bound for the per-step description length of a random walk on the given network.
Thanks to the map equation's foundation in information theory and stochastic processes, it has proven efficient and accurate across various disciplines~\cite{lancichinetti2009pre,Aldecoa2013,10.1371/journal.pone.0154404}.

Over the years, the map equation has been extended in many directions, enabling researchers to analyze increasingly complex data across diverse applications. While powerful, this diversity creates a challenge: users struggle to navigate the variants and select the most suitable approach for their specific needs. Existing map equation tutorials focus on specific applications~\cite{bohlin2014community,edler2017mapequation}, while textbooks and surveys offer high-level descriptions that lack the technical depth needed for practical application~\cite{barabasi2016network, fortunato2016physrep}. This review bridges that gap. We synthesize the map equation's theoretical foundations with practical guidance, helping researchers choose and apply the appropriate variant for their community detection challenges.
Throughout the review, we provide companion Jupyter notebooks with code examples related to the respective section, collected in a GitHub repository at \url{https://github.com/mapequation/infomap-tutorial-notebooks}
and marked with~\faExternalLink~in the text. We use schematic networks for clarity and include a few real-world examples when they offer additional insights.
\section{Mapping Network Flows}

\begin{figure*}[htp!]
    \centering
    \includegraphics[width=\linewidth]{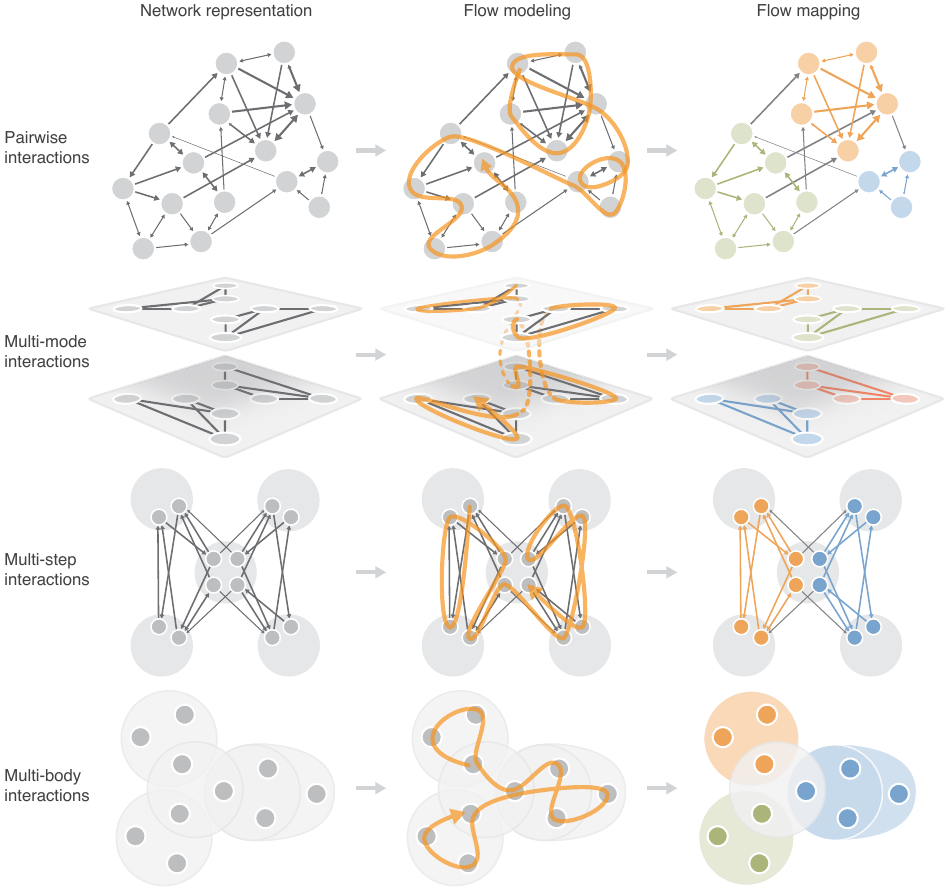}
    \caption{\textbf{Modeling and mapping flow with the map equation framework.}
    Given complex system data, the researcher first selects an appropriate network representation (left column) based on the type of interactions:
    \emph{Pairwise interactions} can be represented with weighted and directed networks, where link strength and direc\-tion capture interaction frequency and orientation.
    \emph{Multi-mode interactions} call for multilayer networks, where nodes are replicated across layers representing different times, contexts, or modes.
    \emph{Multi-step interactions} are captured with memory networks, where physical nodes (large circles) are associated with state nodes (smaller circles) that retain information about interaction sequences.
    \emph{Multi-body interactions} among more than two nodes are naturally represented by hypergraphs, where hyperedges connect multiple nodes simultaneously.
    Next, a random walk model approximates real-world flow (middle column).
    Finally, minimizing the map equation reveals flow modules where a random walker remains for extended periods (right column).
    Because network flows reflect the systems' function, flow modules reveal the systems' functional components. 
    }
    \label{fig:pipeline}
\end{figure*}

Mapping network flows with the map equation framework comprises three steps: network representation, flow modeling, and flow mapping (Fig.~\ref{fig:pipeline}). Each step is adaptable and can be customized to effectively identify flow-based communities in a wide range of scenarios.

\subsection{Network representation}

Complex systems comprise numerous diverse components that interact in intricate ways, making it impossible to comprehend the system's functioning through mere observation.
Network representations seek to schematize the increasingly accessible relational data by abstracting away all but the essential: nodes and links in their most basic form.
The effectiveness of a network representation relies on its ability to explain the complex system through subsequent analysis, including flow-based community detection.

The map equation framework can use a variety of network representations to reliably identify flow-based communities in complex systems (\cref{fig:pipeline}).
In the simplest case, we construct a network with binary links that indicate whether two elements interact.
With more information about interactions, we can consider richer network representations:
Directed and weighted networks to characterize the orientation and strength of interactions.
Multilayer or multiplex networks when the problem requires distinguishing different interaction types or interactions that change over time.
For capturing multi-body interactions, hypergraphs offer a suitable representation.

The research question guides how to best model the characteristics of a real-world complex system.
Can a binary network capture the observed data sufficiently well, or is a weighted directed network required?
Because adding weights and directions improves the model accuracy at a minimal computational cost in the map equation framework, when available we incorporate them with few exceptions.
Will capturing salient temporal features require a multi-step or multi-mode representation?
Are multi-body interactions indispensable for explaining system dynamics?
These types of questions often require testing various network representations, comparing how they influence the flow model and the communities. 

Because any network representation schematizes the raw interaction data, selecting the most suitable network representation involves a delicate balance between simplicity and accuracy:
More intricate network representations prioritize accuracy at the cost of sacrificing conceptual simplicity and computational efficiency.

\subsection{Flow modeling}

Unless the network flows are explicit, we build a flow model on top of the network representation (\cref{fig:pipeline}).
Modeling network flows as Markovian diffusion processes with random walks uses all the information in the network representation with minimal assumptions and can efficiently reveal flow-based communities.
Higher-order generalizations of Markov processes enable incorporating the effects of multi-mode, multi-step, or multi-body interactions.

We first consider pairwise interactions and memoryless diffusion dynamics of a random walk on a network $G = \left(V, E\right)$ with nodes $V$, edges $E$, and $N = \left|V\right|$.
In this setting, the network constrains the random walker's trajectory.
The transition probability $p_{uv}$ that a random walker at node $u$ visits node $v$ in the next step is proportional to the link weight $w_{uv}$ between $u$ and $v$,
\begin{equation}
    p_{uv} = \frac{w_{uv}}{\sum_{v}w_{uv}}, \label{eq:transitionrate}
\end{equation}
and the stationary node visit rates are given by the recursive system of equations
\begin{equation}
    p_{v} = \sum_{u} p_{u} p_{uv}. \label{eq:visitrate}
\end{equation}
In undirected networks, the stationary distribution is proportional to node strength $s$,
\begin{equation}
    p_v = \frac{s_v}{\sum_{u}s_u},
\end{equation}
where $s_v = \sum_{u} w_{uv} = \sum_{u} w_{vu}$ is node $v$'s strength.
That is, the probability that the random walker visits a node is high if the node has high strength.
We refer to random walks to explain the principles behind the map equation but in practice, we do not need to simulate random walks.

However, directed networks can have dangling nodes with no outgoing links or unreachable regions with no incoming links, so the node visit rates depend on where the random walker starts.
In directed networks, random walks are ergodic so that~\cref{eq:visitrate} converges to a unique solution only if the network is strongly connected and the random walk is aperiodic, according to the Perron-Frobenius theorem.
Empirical networks often do not have these properties.
To overcome this issue and ensure a stationary solution in directed networks, we use teleportation and allow the random walker to teleport to a random node independent of the current node at a low rate $\tau$.
Now, stationary node visit rates can be obtained by solving equations
\begin{equation}
    p_{v}^{*} = (1 - \tau) \sum_{u} p_u^{*} p_{uv} + \tau \varepsilon_v \label{eq:teleportation}
\end{equation}
with the power iteration method~\cite{golub2008matrix}.
In \cref{eq:teleportation}, the parameter $\varepsilon_v$ represents the probability that the random walker teleports to the node $v$.
This probability is uniform in the standard teleportation scheme, that is, $\varepsilon_v = \frac{1}{N}$ for all $v$~\cite{gleich-pagerank-beyond-the-web}.

The drawback of this teleportation strategy is that the stationary solution $p_v^{*}$ depends on the choice of $\tau$.
We can apply unrecorded link teleportation, so-called smart teleportation~\cite{lambiotte2012pre} for more robust results.
Instead of assuming a constant probability $\varepsilon_v$ where a random walker teleports to any node irrespective of the network topology, link teleportation selects a link as a teleportation target at random and proportional to its weight.
This leads to a teleportation parameter $\varepsilon_v$ proportional to a node's incoming strength
\begin{equation}
    \varepsilon_v = \frac{s_v^\text{in}}{\sum_{u} s_u^\text{in}},
\end{equation}
where $s_v^\text{in} = \sum_{u} w_{uv}$ denotes the in-strength of node $v$.
Link teleportation reduces the dependence on the parameter $\tau$, however, encoding teleportation steps would affect the detected community structure as this corresponds to introducing artificial links between source and target nodes.
Therefore, we do not record teleportation and only record steps along existing links.
This amounts to performing an extra step without teleportation on a stationary solution for recorded teleportation
\begin{equation}
    p_v^\text{unrec} = \sum_{u} p_u^{*} p_{uv}.
\end{equation}

For any given network representation, all these flow models are unbiased such that a random walker follows links proportional to their weights.
Unbiased random walks are default flow models in the map equation framework. 
Some applications may benefit from a more complex flow model.
For example, a biased flow model where the random walker is guided by node attributes, preferring to visit nodes with similar attributes,
can more accurately model real flow statistics with desirable effects on the flow mapping.

\subsection{Flow mapping}

In modular networks, nodes connect more strongly within communities than between them. As the top row of \cref{fig:pipeline} illustrates, random walkers tend to stay inside communities for extended periods before exiting.

\begin{figure*}[htp!]
    \centering
    \includegraphics[width=0.8\linewidth]{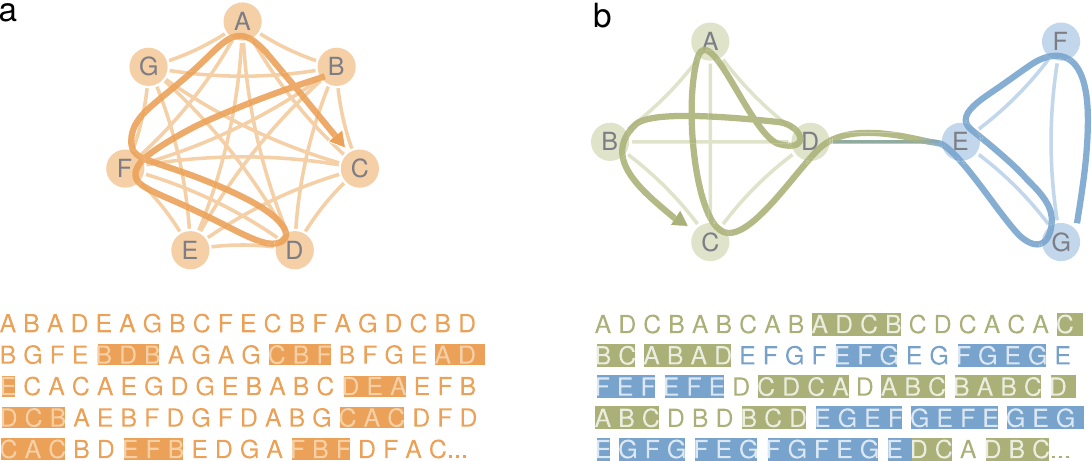}
    \caption{\textbf{Compressing random walk descriptions in modular networks.} 
    (\textbf{a}) In a fully connected network, the random walker visits all nodes uniformly, producing node-visit sequences shown at the bottom. Because repetitions occur randomly, modular compression offers no benefits.
    (\textbf{b}) In a modular network, the walker tends to stay within communities, and the corresponding sequences at the bottom reveal repeated patterns that enable modular compression. Recurring subsequences of length greater than two are highlighted with colored backgrounds.    
    }
    \label{fig:mdl}
\end{figure*}

To capture these modular patterns, we capitalize on the correspondence between detecting regularities in data and compressing those data, formalized by the minimum description length principle~\cite{rissanen1978modeling}.
Compression algorithms aim to represent data compactly by exploiting their regularities. Conversely, we aim to identify regularities in complex systems and measure our success by the compression they enable. The minimum description length principle applied to modular network flows formalizes this duality: the partition that best explains a network's modular flows is the one that most effectively compresses a description of the random walk (\cref{fig:mdl}).

To identify flow modules in a network using the minimum description length principle, we imagine a communication system using efficient codes to transmit the random walk's trajectory. As the random walker moves from node to node, it generates a sequence of node visits. If the network has modular regularities, so will the sequence. By identifying the structure that generates the sequence, we can describe the sequence more compactly with an efficient code~\cite{shannon1948mathematical}. For example, we can exploit frequently visited nodes within modules, much like text compression algorithms exploit repeated patterns. When the code structure mirrors the network's modular organization, the random walk description compresses more effectively.

\begin{figure*}[htp!]
    \centering
    \includegraphics[width=.95\linewidth]{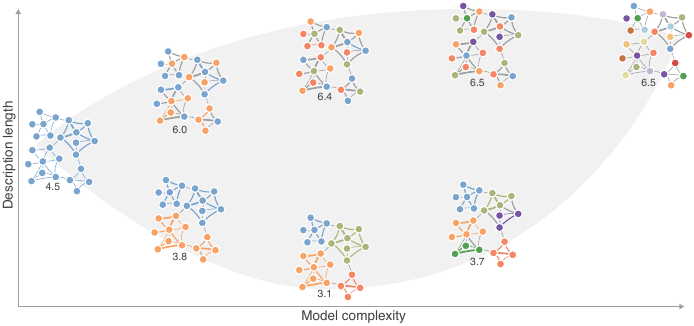}
    \caption{\textbf{Schematic solution landscape of varying model complexity.}
    The number of modules in each network partition defines its model complexity, increasing from left to right. 
    Colors indicate module assignments, and the numbers below solutions approximate the description lengths.
    The solution with the lowest description length balances model complexity and modular regularities.}
    \label{fig:complexity}
\end{figure*}

Applying the minimum description length principle involves finding the model that can explain the data with the shortest possible code, trading off between model complexity and model fit.
A schematic network with four modules illustrates (Fig.~\ref{fig:complexity}): Placing all nodes in a single module minimizes the model complexity with a costly description of the random walker's position among all nodes in the network.
The model maximally underfits the data.
A two-module solution increases the model complexity, requiring specifying movements also between modules. But communicating the random walker's position given the module information is cheaper with fewer nodes to discern.
The shortest possible description length decreases with three and four modules because the cost for model complexity increases slower than the description of the random walker's position given the module information.
Increasing the module complexity beyond four modules leads to overfitting and higher description length in this example. The shortest cost for specifying the position in smaller modules cannot compensate for more complex models. 
The model with one module for each node has the highest model complexity, maximally overfitting the data.

\section{The Map Equation} \label{sec:mapeq}

To employ the minimum description length principle, we conceptually encode the random walker's trajectory with codewords assigned to nodes.
The code's average per-step description length -- we call it \emph{codelength} for short -- measures how well the code exploits regularities in the walker's trajectory.
In practice, we consider not the code itself but the modular compression it enables: the codelength.
Importantly, we can analytically quantify how well different partitions compress expected trajectories without simulating random walks or explicitly assigning codewords.

Grouping nodes into modules enables reusing codewords across modules with one designated codebook per module for a shorter overall codelength.
However, this requires introducing an index-level codebook and module-specific module-exit codewords for describing transitions between modules, increasing the overall codelength.
Finding modules that describe the random walk's regularities well to minimize the codelength means balancing between choosing small modules for efficiently describing intra-module steps and choosing modules such that there are only few links between modules for minimizing the rate at which the random walker takes inter-module steps.

Per Shannon's source coding theorem~\cite{shannon1948mathematical}, the codelength's lower limit when not partitioning nodes into modules is the entropy over the nodes’ stationary visit rates.
Applied to a network's modular structure, the codelength's lower limit is the sum of the module and index-level codebooks' entropies, weighted by the rate at which they are used.

\subsection[The two-level map equation]{\linktonotebook{3.1 The two-level map equation}{The two-level map equation}}

For a partition $\mathsf{M}$ of the nodes $u \in V$ into modules $\mathsf{m} \in \mathsf{M}$, we construct $\left|\mathsf{M}\right|$ module codebooks.
This formulation assumes non-overlapping communities, where each node belongs to exactly one module.
Each node $u$ has a module-dependent codeword, meaning that the same codeword can be reused for nodes in other modules.
For a uniquely decodable code, an additional index codebook encodes transitions into modules.
The index codewords derive from the frequencies of module entries.
The probability $q_{\mathsf{m}\curvearrowleft}$ that a random walker enters module $\mathsf{m}$ is
\begin{equation}
    q_{\mathsf{m}\curvearrowleft} = \sum_{u \notin \mathsf{m}, v \in \mathsf{m}} p_{u} p_{uv}.
\end{equation}
Besides codewords for communicating which node a random walker visits, each module codebook requires a designated exit codeword for communicating when a random walker leaves the module.
The probability $q_{\mathsf{m}\curvearrowright}$ of leaving module $\mathsf{m}$ is
\begin{equation}
    q_{\mathsf{m}\curvearrowright} = \sum_{u \in \mathsf{m}, v \notin \mathsf{m}} p_{u} p_{uv}.
\end{equation}

To define the map equation, we apply Shannon’s source coding theorem to the index codebook and each module codebook.
The codelength for describing movements within module $\mathsf{m}$ is
\begin{equation}
    H(P_\mathsf{m}) = -\frac{q_{\mathsf{m}\curvearrowright}}{p_\mathsf{m}^{\circlearrowright}} \log_2 \frac{q_{\mathsf{m}\curvearrowright}}{p_\mathsf{m}^{\circlearrowright}} - \sum_{u \in \mathsf{m}} \frac{p_u}{p_\mathsf{m}^{\circlearrowright}} \log_2 \frac{p_u}{p_\mathsf{m}^{\circlearrowright}},
\end{equation}
where $p_\mathsf{m}^{\circlearrowright} = q_{\mathsf{m}\curvearrowright} + \sum_{u \in \mathsf{m}} p_u$ is the total use rate of module $\mathsf{m}$'s codebook.
Similarly, for the index codebook, the codelength for describing random walker transitions between modules is
\begin{equation}
    H(Q) = - \sum_{\mathsf{m} \in \mathsf{M}} \frac{q_{\mathsf{m}\curvearrowleft}}{q_{\curvearrowleft}} \log_2 \frac{q_{\mathsf{m}\curvearrowleft}}{q_{\curvearrowleft}},
\end{equation}
where $q_{\curvearrowleft} = \sum_{\mathsf{m} \in \mathsf{M}} q_{\mathsf{m}\curvearrowleft}$.

The sum of codelengths for all codebooks weighted by their use rate gives the map equation
\begin{equation}
    L\left(\mathsf{M}\right) = q_{\curvearrowleft} H(Q) + \sum_{\mathsf{m} \in \mathsf{M}} p_\mathsf{m}^{\circlearrowright} H(P_\mathsf{m}). \label{eq:mapeq}
\end{equation}
With an increasing number of modules, the codelength of the module codebooks decreases while the codelength of the index codebook increases.
The network partition that best captures the modular network flows maximally compresses the description length defined by the map equation -- minimizing the map equation over all possible partitions gives the optimal partition.

\subsection{The multilevel map equation}
\begin{figure}[ht]
    \centering
    \hfill
    \begin{minipage}[t]{0.3\linewidth}
    \includegraphics[width=\linewidth]{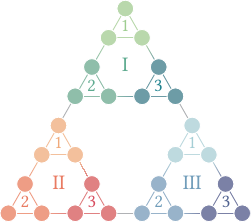}
    \end{minipage}
    \hfill
    \begin{minipage}[t]{0.62\linewidth}
    \vspace*{-7.5ex}
    \captionsetup{singlelinecheck=off, skip=0pt, justification=raggedright}
    \caption{\textbf{Schematic triangle network with a multilevel partition.}
    Each small triangle (1, 2, 3) is a sub-module, organized in groups of three triangles in each top-level super-module (I, II, III).}
    \label{fig:multilevel-map-equation}
    \end{minipage}
\end{figure}
Complex systems often have a hierarchical -- \emph{multilevel} -- organization consisting of nested modules at different levels (Fig.~\ref{fig:multilevel-map-equation}).
We generalize the map equation to describe multilevel structures by applying it to each module $\mathsf{m} \in \mathsf{M}$ recursively~\cite{rosvall2011multilevelmapeq, edler2017mapequation},

\begin{minipage}{\linewidth}
\begin{align}
  L(\mathsf{M}) = q_{\curvearrowleft} H(Q) + \sum_{\mathsf{m} \in \mathsf{M}} L(\mathsf{m}).\label{eq:recursivemapeq}
\end{align}
\end{minipage}

Each module $\mathsf{m}$ contains either (i) a set of submodules or (ii) a set of nodes:
\begin{enumerate}
  \item[i.] For a module $\mathsf{m}$ with submodules $\mathsf{m}' \in \mathsf{m}$, the contribution to the overall codelength is
  \begin{align}
    L\left(\mathsf{m}\right) = q_{\mathsf{m}}^\circlearrowright H\left(Q_{\mathsf{m}}\right) + \sum_{\mathsf{m}' \in \mathsf{m}} L\left(\mathsf{m}'\right),
  \end{align}
  that is, we treat each module $\mathsf{m}$ with submodules the same way as the top-level module $\mathsf{M}$, with one addition: we need to consider module exits from $\mathsf{m}$, thus defining its codebook usage rate
  \begin{equation}
    q_{\mathsf{m}}^{\circlearrowright} = q_{\mathsf{m}\curvearrowright} + \sum_{\mathsf{m}' \in \mathsf{m}} q_{\mathsf{m}'\curvearrowleft}
  \end{equation}
  and
  \begin{equation}
      H\left(Q_\mathsf{m}\right) = - \frac{q_{\mathsf{m}\curvearrowright}}{q_\mathsf{m}^\circlearrowright} \log_2 \frac{q_{\mathsf{m}\curvearrowright}}{q_\mathsf{m}^\circlearrowright} - \sum_{\mathsf{m}' \in \mathsf{m}} \frac{q_{\mathsf{m}'\curvearrowleft}}{q_\mathsf{m}^\circlearrowright} \log_2 \frac{q_{\mathsf{m}'\curvearrowleft}}{q_\mathsf{m}^\circlearrowright}.
  \end{equation}
  
  \item[ii.] The codelength contribution of a module $\mathsf{m}$ that contains nodes but no further submodules is
  \begin{equation}
    L\left(\mathsf{m}\right) = p_\mathsf{m}^\circlearrowright H \left(P_\mathsf{m}\right),
  \end{equation}
  where
  \begin{equation}
    p_\mathsf{m}^\circlearrowright = q_{\mathsf{m}\curvearrowright} + \sum_{u \in \mathsf{m}} p_u \label{eq:usemno}
  \end{equation}
  and
  \begin{equation}
      H(P_\mathsf{m}) = 
       -\frac{q_{\mathsf{m}\curvearrowright}}{p_\mathsf{m}^\circlearrowright} \log_2 \frac{q_{\mathsf{m}\curvearrowright}}{p_\mathsf{m}^\circlearrowright}
       - \sum_{u \in \mathsf{m}} \frac{p_u}{p_\mathsf{m}^\circlearrowright} \log_2 \frac{p_u}{p_\mathsf{m}^\circlearrowright}. \label{eq:Hijl}
  \end{equation}
\end{enumerate}

\subsection{Challenges and remedies}

\paragraph{\textbf{Resolution limit.}}
Finding modules in large networks can be affected by a resolution limit that prevents detecting small modules when large modules are present.
Small but functionally meaningful modules may be merged into larger ones.

For the map equation, there is an analytical estimate that relates the number of links between modules to the map equation's resolution limit,
\begin{equation}
    \frac{4^{l_\mathsf{m}}}{l_\mathsf{m} + 1} \lesssim C
\end{equation}
where $l_\mathsf{m}$ is the number of internal links in module $\mathsf{m}$, and $C$ is the cut size, that is, the number of links between modules~\cite{kawamoto2015pre}.
However, compared to other widely used community detection methods, such as Modularity, whose smallest detectable module scales as $\mathcal{O}(\sqrt{E})$ with the total number of links in the network~\cite{fortunato2007pnas}, and the stochastic block model, which scales as $\mathcal{O}(\sqrt{N})$ with the total number of nodes~\cite{peixoto2013prl}, the map equation is less affected by resolution limit.

\paragraph{\textbf{Field-of-view limitations.}}
\begin{wrapfigure}{r}{.45\linewidth}
    \centering
    \includegraphics[width=\linewidth]{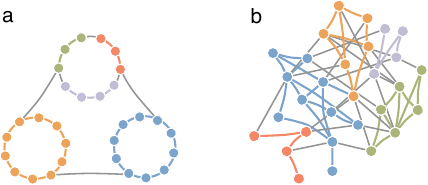}
    \caption{\textbf{Field-of-view limitations.} Over-partitioning can occur in networks with (a) constrained structure, or (b) random structure.
    Node colors show module assignments.}
    \label{fig:field_of_view_limit}
\end{wrapfigure}
Dual to the resolution limit, which can prevent detecting modules below an effective size, methods can have an upper limit in the effective size they can detect, a so-called field-of-view limit.
While the map equation is less susceptible to the resolution limit than other community detection methods, the field-of-view limit can affect its performance in some networks.

The map equation's flow-based approach implicitly assumes that modules are assortative structures where nodes with similar functions are densely connected.
However, this assumption may not be valid in networks with constrained structures.
For example, geographically embedded networks such as power grids or transportation networks typically contain sparse regions.
The map equation tends to over-partition these regions into smaller communities~\cite{schaub2012plosone,schaub2012pre}.
We illustrate a realization of this problem in \cref{fig:field_of_view_limit}a.

Over-partitioning can also occur in random networks created without an explicit modular structure.
The map equation aims to describe the network's structure concisely but without assuming any generative process.
Consequently, the map equation identifies modular structure in random networks whose density is below a certain threshold, simply because such networks contain subgroups of nodes with low external link density~\cite{lancichinetti2009pre} (\cref{fig:field_of_view_limit}b).

\paragraph{\textbf{Markov time scaling.}}
By integrating Markov time scaling with the map equation to adjust its resolution, we can overcome the field-of-view limitations~\cite{kheirkhahzadeh2016pre}.
The standard map equation encodes the random walker's position in the network after every transition; thus, the default Markov time is $t=1$.
However, we can generalize the map equation to Markov times other than one by scaling the transition rates in \cref{eq:transitionrate} with the Markov time $t$ 
\begin{equation}
    p_{uv}\left(t\right) = t p_{uv},
\end{equation}
and if $t<1$, we additionally account for
\begin{equation}
    p_{uu}\left(t\right) = \left(1-t\right) + t p_{uu}.
\end{equation}
This scaling approximates continuous-time transition probabilities between nodes $u$ and $v$, which grow linearly with time. If we let the walker take $t$ steps before encoding its position, it transitions between nodes approximately $t$ times more than if it only took one step. While the stationary distribution of node visits remains unchanged, the transition rates between nodes increase in proportion to $t$.
For $t<1$, the random walker visits the same node several times because it moves slower, resulting in smaller modules.
For $t>1$, the walker moves faster and can take more than one step before we encode such that not every node along the trajectory is encoded, favoring larger modules.

\paragraph{\textbf{Variable Markov time scaling.}}
Variable Markov time~\cite{edler2022variable} relaxes the constraints of a global Markov time parameter: choosing between detecting large-scale structures above the field-of-view limit or highlighting small dense structures.
The basic idea is to adjust the Markov time dynamically based on the random walker's current neighborhood in the network, allowing it to move faster in sparse areas and slower in dense areas, reducing the effects of both the field-of-view limit and the resolution limit simultaneously.
Adjusting the random walker's Markov time dynamically enables capturing a broader range of modular flow patterns within a single map.

To scale the Markov time dynamically, we adjust the flow between nodes with a node-local Markov time~$t_u$ that is inversely proportional to the flow of node $u$.
As a baseline, we use the densest part of the network and keep the minimum Markov time at~1 to avoid the resolution limit.
We define the variable Markov time $t_u$ for node $u$ as the ratio between the maximum flow over all nodes, $p_\text{max}$, and the local flow~$p_u$,
\begin{equation}
    t_u = \frac{p_\text{max}\,k_\text{tot}}{\min(1, p_u k_\text{tot})},
    \label{eqn:variable-markov-unskewed}
\end{equation}
where $k_\text{tot} = \sum_u k_u$ is the total node degree over all nodes in the network.
However, in real-world networks where most nodes typically have few links and a few nodes have many links, this definition of variable Markov time is sensitive to the degree of the strongest node.
We can address this issue by taking the logarithm with base 2 of $p_\text{max}$  and $p_u$ instead to avoid small Markov times, resulting in a variable Markov time where the random walker moves with a constant entropy rate, 
\begin{equation}
    t_u = \frac{\log_2 \left(p_\text{max}\,k_\text{tot}\right)}{\min\left(1, \log_2 \left(p_u\,k_\text{tot}\right)\right)}.
    \label{eqn:variable-markov-skewed}
\end{equation}

For directed links, we scale the flow with the Markov time of the source nodes.
To keep the minimum Markov time at $1$ for undirected links, we scale the flow with the minimum Markov time $t_{uv} = \min(t_u, t_v)$ between nodes~$u$ and~$v$.
Typically, networks with less skewed degree distribution can benefit from using \cref{eqn:variable-markov-unskewed} to avoid the field-of-view limit.

\begin{figure*}[htp!]
    \centering
    \includegraphics[width=\linewidth]{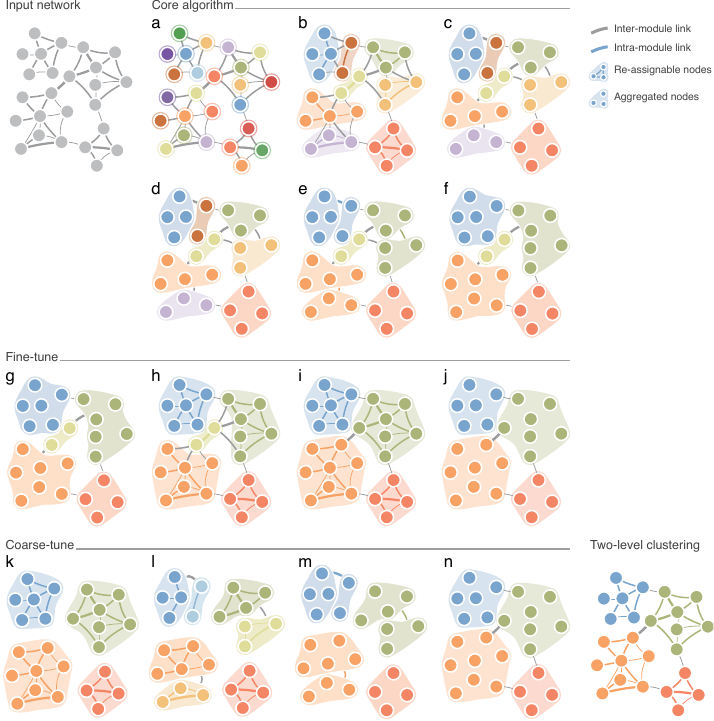}
    \caption{\textbf{Schematic illustration of Infomap's search algorithm for two-level partitions.}
    Infomap uses a greedy stochastic search algorithm to optimize the map equation over modular regularities.
    The algorithm combines multiple strategies to avoid local minima. It begins by placing each node in its own module (a), then iteratively moves nodes to neighboring or new modules if such moves reduce the codelength (b--c).
    Next, it moves the identified modules together as a unit and repeats the same logic at a higher level (d--f). This recursive grouping repeats as long as the total codelength decreases beyond a minimal threshold.
    To refine module assignments at multiple scales, Infomap alternates between fine-tuning (g--j) and coarse-tuning (k--n).
    During fine-tuning, it revisits node-level assignments (b--c) starting from the current modular structure, again moving individual nodes between modules to reduce the codelength (g--j).
    During coarse-tuning, Infomap first partitions each module independently to identify sub-modular groups (l),
    reassigns these groups to the global modular structure (m), and moves them between modules using the higher-level logic (d--f), continuing as long as the codelength decreases beyond a minimal threshold (n).
    }
    \label{fig:search-algorithm}
\end{figure*}

\section{Infomap} \label{sec:infomap}

Infomap is a greedy stochastic search algorithm designed to minimize the map equation and detect two-level and multilevel flow communities in networks~\cite{rosvall2011multilevelmapeq,bohlin2014community}.
Since community detection is an NP-hard combinatorial optimization problem, Infomap cannot guarantee to find the map equation's global minimum; it uses iterative and recursive optimization heuristics to avoid local minima and is known to achieve good results in practice \cite{lancichinetti2009pre,Aldecoa2013,10.1371/journal.pone.0154404}.
The Infomap search algorithm is inspired by the Louvain algorithm for modularity maximization~\cite{blondel_fast_2008} but uses additional fine-tuning and coarse-tuning steps, similar to how the Leiden algorithm later refined Louvain~\cite{traag_louvain_2019}.
When expressed in differentiable tensor form with soft cluster assignments, the map equation can also be optimized with graph neural networks~\cite{blöcker2023map}.
Infomap is available as an open-source software package~\cite{mapequation2022software}.

Infomap runs in two phases: the first phase creates a two-level partition, and the second phase creates a multilevel partition.
Depending on the use case, Infomap can stop either after the two-level phase and output a two-level partition or after the multilevel phase for a multilevel partition if it improves the two-level solution.

\subsection[The two-level phase]{\linktonotebook{4.1 The Two-level Phase}{The two-level phase}}
The two-level phase consists of two stages:

In stage 1, first, Infomap assigns each node to its own module, creating a trivial partition with high codelength as a starting point (\cref{fig:search-algorithm}a).
Such a partition of singletons places all links between modules -- assuming the network has no self-loops -- and forces the random walker to switch modules with every step, inefficiently compressing the random walker's trajectory.
Second, Infomap updates the partition iteratively and considers moving nodes in random order to either of their neighboring modules or a new singleton module (\cref{fig:search-algorithm}b).
To avoid re-checking nodes with stable module assignments, Infomap initially marks all nodes as candidates for moving.
Then, as long as there are candidates left, Infomap randomly chooses a candidate and tries to move it to reduce the codelength by some adjustable minimum amount.
If there are several possible moves for a node, Infomap chooses the one that reduces the codelength the most.
When Infomap moves a node, it marks that node and all its neighbors as candidates for moving.
A node that has no improving move is unmarked.
When no candidate nodes are left, Infomap stops moving individual nodes (\cref{fig:search-algorithm}c).
Third, Infomap repeats the second step, now moving the identified modules: nodes that belong to the same module are moved into or out of other modules together as a group.
This step repeats with larger and larger modules as long as the codelength decreases (\cref{fig:search-algorithm}d--f).
We call stage 1 Infomap's \emph{core algorithm}.

In stage 2, Infomap tunes the partition to avoid local minima by alternating between fine-tuning and coarse-tuning steps.
Fine-tuning and coarse-tuning follow the same principle and use the core algorithm, but they consider different objects to move.
For fine-tuning, Infomap uses the core algorithm to move individual nodes to improve the codelength (\cref{fig:search-algorithm}g--j).
For coarse-tuning, Infomap partitions all modules individually into sub-modules using the core algorithm and then moves these sub-modules to reduce the codelength (\cref{fig:search-algorithm}k--n).
Infomap fine- and coarse-tunes at least once and stops when it cannot improve the codelength by some minimum amount:
either by an absolute value or a relative value determined from the codelength of the one-level partition, both of which can be adjusted.

\subsection{The multilevel phase}
The multilevel phase aims to reduce the codelength by adding further index levels to a two-level partition.
It contains two stages.

In stage 1, Infomap compresses inter-module transitions by first aggregating the network at the module level.
This creates a network where nodes represent the previous modules, and inter-module links are merged.
Second, Infomap uses the two-level algorithm to partition the aggregated network.
The resulting two-level partition comprises a three-level partition when interpreted in the context of the network before aggregation.
Infomap repeats stage 1 as long as aggregating and partitioning the network and adding one more index level per iteration yields a non-trivial solution. 

In stage 2, Infomap partitions the modules at the highest level of the multi-level partition recursively.
Since the lower levels in the partition were designed from a perspective that did not consider the higher levels, they may be suboptimal and suffer from resolution issues.
Moreover, the resulting partition is balanced, and all modules have the same depth, which is generally not the case in real-world networks.
To overcome this, Infomap keeps the modules at the highest level but forgets their submodules to re-partition each module independently, using the full algorithm, that is, the two-level phase followed by the multi-level phase.

\subsection[Solution landscape]{\linktonotebook{4.3 Solution Landscape}{Solution landscape}}

Even though Infomap is designed to avoid converging to a local minimum, the non-convex properties of the map equation can be a problem in practice.
Moreover, local minima can have a codelength close to the global minimum, which can be an inherent property of the studied system, or be caused by noisy data or sensitivity to model parameters.

To study the map equation's solution landscape, we visualize the detected minima~\cite{calatayud2019solution}.
We run Infomap repeatedly and with different seeds to generate different partitions and cluster them to identify the partition clusters in the map equation's solution landscape.
We consider the solution landscape complete when new partitions fall within already identified partition clusters with a high probability.
Reaching a complete landscape also answers how many times we need to run Infomap to find all relevant solutions.
To measure distances between partitions, we use a Jaccard-based metric that allows using two-level or multilevel partitions.
By varying the cluster distance threshold we can explore the solution landscape at different scales.

\begin{wrapfigure}[24]{r}{0.5\linewidth}
    \centering
    \includegraphics[width=\linewidth]{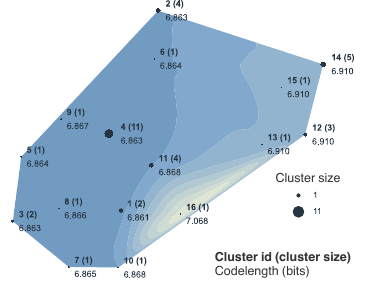}
    \caption{\textbf{The solution landscape of the jazz network.} Visualization of clusters of partitions and the codelength of the best partition in each cluster.
    Cluster 1 has two partitions and the shortest codelength. 
    Clusters 2--4 have a slightly higher codelength and comprise 17 partitions.
    Differences between partitions of similar quality can reveal important aspects of the underlying data.
    Color indicates the solution quality, with light green for long codelengths and dark blue for short codelengths. The UMAP embedding preserves local relationships; similar solutions appear close to each other in the solution landscape.}
    \label{fig:solution-landscape}
\end{wrapfigure}

\Cref{fig:solution-landscape} shows a two-dimensional embedding of the jazz network's~\cite{Gleiser2003jazz} solution landscape, created with UMAP~\cite{McInnes2018UMAP}.
The jazz network's solution landscape has several local minima with codelengths close to the global minimum.
If these alternative partitions differ in any aspect important to the researcher -- compared to each other or the global minimum -- they can reveal important aspects of the underlying data.
In contrast, the solution landscape of the Zachary karate club network~\cite{Zachary} has only one partition, that is, Infomap always converges to the same solution.

\section{The map equation for higher-order networks} \label{sec:higherorder}

Higher-order network models can provide a more accurate representation of complex systems by capturing dynamics beyond pairwise interactions~\cite{battiston_networks_2020,torres_why_2021}.
These networks have found applications in various fields where conventional network models may fall short.
For example, conventional networks may struggle to represent protein--protein interactions, where three or more proteins can form complexes.
Similarly, they may not adequately capture the dynamics of social networks, where the information flow depends on its source.

To address these limitations, the higher-order map equation framework introduces so-called state nodes to represent higher-order dependencies and generalizes Markovian dynamics from first to higher order.
State nodes can reflect various aspects of the system, such as different layers in multilayer networks or previously visited physical nodes in memory networks.
Physical nodes correspond to regular nodes in first-order networks and represent the interacting entities.

We use the map equation to partition nodes into communities that minimize the description of random walker movements between state nodes. 
This generalization does not impact the index codebook but necessitates modifications to the module codebooks.
When multiple state nodes of the same physical node are assigned to the same module, they share the same codeword to ensure they represent the same object. 
At the same time, state nodes of the same physical node assigned to different modules allow identifying overlapping modules.

We discuss various higher-order models focusing on memory networks, multilayer networks, temporal networks, and hypergraphs.
While the map equation directly supports multilayer and memory networks, temporal networks and hypergraphs require additional modeling choices.
We build on the memory and multilayer map equation and incorporate additional constraints into the random-walk model to capture the intricacies of temporal and multi-body interactions.

\subsection[Memory networks]{\linktonotebook{5.1 Memory Networks}{Memory networks}}
\begin{figure}[hpt!]
    \centering
    \includegraphics[width=.7\linewidth]{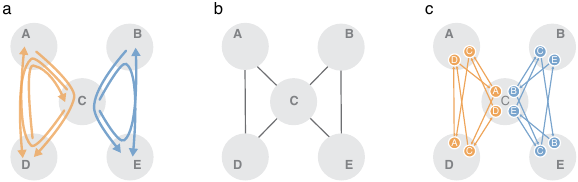}
    \caption{\textbf{Two-step paths represented as a memory network.}
    Two-step paths (a) are realized as a first-order network in (b) and a second-order memory network using state nodes in (c).
    Each state node represents the memory of the previous step: The path $\mathsf{A-C-D}$ is represented by the state node $\mathsf{A}$ in physical node $\mathsf{C}$ connected to the state node $\mathsf{C}$ in $\mathsf{D}$.
    With more pathway data, each physical node may comprise multiple state nodes.
    The memory network can be partitioned into overlapping modules by assigning the state nodes belonging to the same physical node to different modules.}
    \label{fig:schematic-higher-dynamics}
\end{figure}

In a $k$-th order memory network, the random walker's next step depends on the $k-1$ previously visited nodes
\begin{equation}
    P_{uv} = P(v | u, u_{-1}, \dots, u_{-k+1}),
\end{equation}
where $u$ is the current node, $u_{-1}$ the previous node, and $u_{-k}$ the node visited $k$ steps ago.
Modeling with memory networks enables representing flow phenomena such as high return probabilities.
Memory networks have been used successfully in various domains, including modeling citation pathways through multidisciplinary journals and traffic flow through transit airports~\cite{rosvall_memory_2014,persson2016maps}.

We illustrate the approach using flows from two domains through a central node~$\mathsf{C}$ (\cref{fig:schematic-higher-dynamics}a).
Conventional first-order networks cannot model the flow dependency on its origin, resulting in undesired mixing (\cref{fig:schematic-higher-dynamics}b).
In contrast, a second-order memory network overcomes this limitation and encodes the history of previously visited physical nodes in state nodes:
Here, the state node $\mathsf{A}$ in physical node $\mathsf{C}$ represents the transition from $\mathsf{A}$ to $\mathsf{C}$ (\cref{fig:schematic-higher-dynamics}c). 
Using state nodes allows us to separate between flows that mix in physical node $\mathsf{C}$ and identify two overlapping modules.

\paragraph{\textbf{Memory networks from first-order data.}}

To obtain a stationary flow distribution for state nodes in a memory network, we require sequence data; ideally, empirical multi-step data.
However, in cases where such data are unavailable, we can model higher-order data on top of first-order networks to reveal overlapping modules.
This approach offers a valuable alternative for identifying patterns in complex systems with higher-order dependencies, even in the absence of empirical multi-step data.

For example, we can use the second-order random walk model known as \textit{node2vec}~\cite{grover_node2vec_2016} to create state nodes.
In this model, the return parameter~$p$ controls the probability that flows return to where they came from, while the in-out parameter~$q$ controls the probability that flows stay in the vicinity or move further away from where they came from.
By tuning $p$ and~$q$, flows can be directed to stay more local to where they came from, reinforcing triangles~\cite{holmgren2023biasedwalks}.
To avoid representing every possible second-order pathway as state nodes, we can derive a compact memory network by lumping state nodes based on minimum information loss~\cite{holmgren2023biasedwalks}.

\subsection[Multilayer networks]{\linktonotebook{5.2 Multilayer Networks}{Multilayer networks}}

\begin{figure*}[thp!]
    \centering
    \includegraphics[width=.95\linewidth]{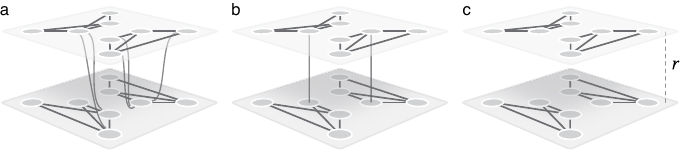}
    \caption{Multilayer network models with (a) explicit multilayer links, (b) intra- and inter-layer links, and (c) intra-layer links and inter-layer relax rate.}
    \label{fig:multilayer-models}
\end{figure*}

\newcommand{\Multilayer}{\texttt{*Multilayer}}
\newcommand{\Intra}{\texttt{*Intra}}
\newcommand{\Inter}{\texttt{*Inter}}

Real-world systems are often characterized by flows that can be stratified across different layers.
For example, individuals may use different modes of transport to commute between cities or communicate with each other using various messaging or social media platforms.
To model such multilayer flows, we utilize multilayer networks where each physical node can exist in many layers with different link structures. 
A group of nodes with similar connectivity patterns across different layers should be assigned to the same community.
However, if the connectivity patterns of a physical node vary across the layers, it should be assigned to multiple communities.

The map equation interprets a multilayer network with $l$ layers as an $l$-th-order memory network where state nodes assigned to a physical node correspond to different layers \cite{de_domenico_identifying_2015}.
When connecting multilayer state nodes, we distinguish two types of links: intra-layer links, which connect nodes within the same layer, and inter-layer links, which connect nodes across different layers.
Depending on the input data, there are three ways to define inter-layer links.

\paragraph{\textbf{Explicit multilayer links.}}
If the data provide explicit information about the link weight $w_{u^A v^B}$ between node $u$ in layer $A$ and node $v$ in layer $B$ (\cref{fig:multilayer-models}a), we can directly obtain a stationary flow distribution for the state nodes.
The transition probability between any two state nodes $u^A$ and $v^B$ is
\begin{equation}
    p_{u^A v^B} = \frac{w_{u^A v^B}}{\sum_{v^B} w_{u^A v^B}}. \label{eq:transitionrate_multilayer}
\end{equation}

\paragraph{\textbf{Intra- and inter-layer links.}}
In some cases, the data lack explicit multilayer links between all pairs of state nodes.
Instead, it may provide explicit links within layers, and inter-layer link weights $\upsilon_{u^A u^B}$, which describe the coupling between physical node $u$ in layers $A$ and $B$ when $A \neq B$ (\cref{fig:multilayer-models}b).
In this model, the random walker moves between layers proportional to inter-layer link weight, but we do not encode this movement.
The map equation expands the inter-layer links to multilayer links, and the transition probability is
\begin{equation}
    p_{u^A v^B} = \frac{\upsilon_{u^A u^B}}{\sum_B \upsilon_{u^A u^B}} \frac{w_{u^B v^B}}{\sum_{v} w_{u^B v^B}}, \label{eq:transitionrate_intrainter}
\end{equation}
where $\upsilon_{u^A u^A} = \sum_{v} w_{u^A v^A}$.

\paragraph{\textbf{Intra-layer links and inter-layer relax-rate.}}
In most cases, empirical inter-layer links are not provided.
To relax the layer constraints and allow the random walker to move without inter-layer links, the map equation introduces a relax rate parameter $r$ (\cref{fig:multilayer-models}c).
In this model, the random walker follows intra-layer links with a probability $1-r$ and moves between layers with a probability $r$.
With the Kronecker delta $\delta$, the transition probabilities become
\begin{equation}
    p_{u^A v^B} = \left(1-r\right)\delta_{AB}\frac{w_{u^A v^B}}{\sum_{v^B}w_{u^A v^B}} + r\frac{w_{u^A v^B}}{\sum_A\sum_{v^B}w_{u^A v^B}}. \label{eq:transitionrate_relaxrate}
\end{equation}

\subsection[Modeling temporal data]{\linktonotebook{5.3 Modeling Temporal Data}{Modeling temporal data}}
\begin{wrapfigure}{r}{.6\linewidth}
    \centering
    \includegraphics[width=\linewidth]{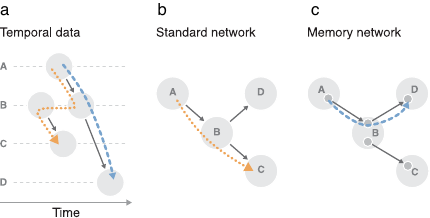}
    \caption{\textbf{Standard network models destroy temporal information.}
    (a) Drawing interactions along a time axis shows that the orange random walker moves back in time, thus breaking causality.
    The blue random walker guided by state nodes obeys causality.
    (b) Conventional first-order network models allow the orange random-walker step because they do not model temporal dependencies.
    (c) Memory networks constrain temporal flows according to available data.
    }
    \label{fig:temporal-network}
\end{wrapfigure}

Memory and multilayer networks offer two alternatives to capture higher-order dependencies in temporal network data.
Memory networks capture causal paths that standard networks wash out~\cite{lambiotte2019networks}.
For example, $\mathsf{A}$ can only causally influence $\mathsf{C}$ through $\mathsf{B}$ if $\mathsf{A}$ influences $\mathsf{B}$ before $\mathsf{B}$ influences $\mathsf{C}$, but conventional first-order networks discard temporal ordering and distort essential flow dynamics critical for unveiling system functions (\cref{fig:temporal-network}).
In contrast, state nodes in memory networks preserve temporal ordering.

When dealing with network snapshots across different time intervals rather than pathway data, we model the data with a temporal multilayer network.
Such a model allows us to represent the network's evolution over time by organizing snapshots into distinct layers and enables analyzing, for example, temporal communities in ecological systems and fossil record data~\cite{farage_identifying_2021, rojas2021multiscale}.
In these systems, inter-layer links are typically absent; instead, a relaxation rate models random-walker transitions between layers.
To control temporal ordering in multilayer networks explicitly, the map equation can apply layer constraints to restrict the random walker to jump only to layers within a specific temporal distance~\cite{Aslak2018}.

Furthermore, an adapted version of the map equation captures intermittent communities that emerge and dissolve repeatedly and independently from other communities in the network~\cite{Aslak2018}.
Because the multilayer map equation couples entire layers, it cannot accurately capture flow persistence within groups of nodes whose existence is time-dependent, making it difficult to identify intermittent communities.
To address this scenario, we use node-level layer coupling:
Node-level layer coupling estimates the coupling strength $\upsilon_{u^A u^B}$ between node $u$ in layers $A$ and $B$ based on how similar $u$'s connectivity patterns are in these two layers, quantified using the Jensen-Shannon divergence (JSD)
\begin{equation}
    \upsilon_{u^A u^B} = 1 - JSD(p_{u^A}, p_{u^B}),
\end{equation}
where $p_{u^A} = (p_{u^Av_1^A}, \dots, p_{u^Av_N^A})$ represents the intra-layer probabilities that a random walker steps from node $u$ in layer $A$ to other nodes $v$ in layer $A$.

\subsection[Modeling multi-body interactions]{\linktonotebook{5.4 Modeling Multi-body Interactions}{Modeling multi-body interactions}}
We can identify modules in hypergraphs by representing random walks on hypergraphs as walks on conventional or multilayer networks.
However, different systems and research questions necessitate different random-walk and hypergraph models.
Hyperedges can have weights, and nodes can have hyperedge-dependent weights~\cite{chitra_random_2019}.
Random walks can be lazy, allowing multiple consecutive visits to the same node, or non-lazy, forcing it to move on~\cite{carletti_random_2020}.
Additionally, these random-walk models can be represented using various network types, such as bipartite, unipartite, or multilayer networks.
What network representation to choose depends on the research question, as each has different advantages.

\begin{figure}[htp!]
    \centering
    \includegraphics[width=0.6\linewidth]{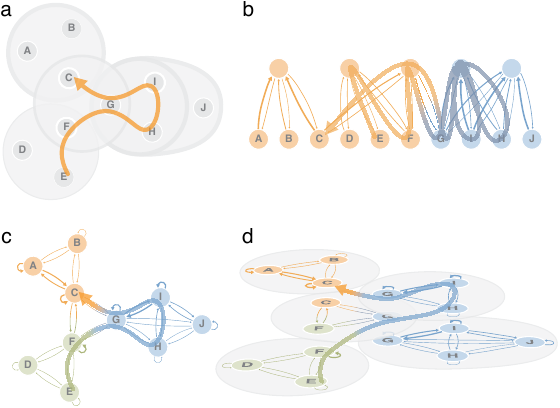}
    \caption{\textbf{Schematic hypergraph represented with three types of networks.}
    (a) The schematic hypergraph with weighted hyperedges and hyperedge-dependent node weights. White circles labeled from A to J represent nodes, and large circles represent hyperedges incident to the nodes in each circle. 
    Node and hyperedge borders indicate weight.
    A random walk depicted with an arrow on the schematic hypergraph represented on: (b) a bipartite network where the unlabelled nodes represent the hyperedges, (c) a unipartite network, and (d) a multilayer network with grey circles defining each layer.
    Node colors indicate optimized module assignments, links' thicknesses are proportional to the random walk's transition rates.  The figure is adapted from Ref.~\cite{eriksson_how_2021}, licensed under CC BY 4.0.}
    \label{fig:hypergraph-reps}
\end{figure}

We model flows on hypergraphs with random walks, using
hypergraphs with nodes $V$, hyperedges $E$ with weights $\omega$, and hyperedge-dependent node weights $\gamma$.
Each hyperedge~$e$ has a weight $\omega(e)$.
Each node~$u$ has a weight $\gamma_e(u)$ for each hyperedge $e$ incident to $u$.
We denote node $u$'s total incident hyperedge weight $d(u) = \sum_{e \in E(u)}\omega(e)$ and hyperedge~$e$'s total node weight $\delta(e) = \sum_{u \in e}\gamma_e(u)$.
With these weights, a lazy random walker moves from node~$u$ to~$v$ in three stages (\cref{fig:hypergraph-reps})\cite{chitra_random_2019}
\begin{enumerate}
    \item Picking hyperedge $e$ among node $u$'s hyperedges $E(u)$ with probability $\frac{\omega(e)}{d(u)}$.
    \item Picking one of hyperedge $e$'s nodes $v$ with probability  $\frac{\gamma_e(v)}{\delta(e)}$.
    \item Moving to node $v$.
\end{enumerate} 
Variations include non-lazy walks, which never visit the same node twice in a row, with a modified second stage
\begin{enumerate}\setcounter{enumi}{1}
\item[(2b)] Picking one of hyperedge $e$'s nodes $v \ne u$ with probability  $\frac{\gamma_e(v)}{\delta(e)-\gamma_e(u)}$.
\end{enumerate} 

\paragraph{\textbf{Bipartite.}}
Bipartite networks offer the most direct representation of the basic three-stage random-walk process.
We represent the hyperedges with \textit{hyperedge nodes}, and the three stages become a two-step walk between the nodes at the bottom and the hyperedge nodes at the top in \cref{fig:hypergraph-reps}b.
First, a step from a node~$u$ to a hyperedge~node~$e$,
\begin{align}
    P_{ue} = \frac{\omega(e)}{d(u)},
\end{align}
and then a step from the hyperedge node to a node~$v$,
\begin{align}
    P_{ev} = \frac{\gamma_e(v)}{\delta(e)}.
\end{align}
For non-lazy walks, we let each incoming link to a hyperedge node connect to a state node with out-links to all of the hyperedge's nodes except the incoming link's source node.

\paragraph{\textbf{Unipartite.}}
To represent the random walk on a unipartite network, we project the three-stage random-walk process down to a one-step process between the nodes and describe it with the transition rate matrix
\begin{align}
    P_{uv} = \hspace{-1em}\sum_{e \in E(u,v)}\hspace{-1em} P_{ue}P_{ev} = \hspace{-1em}\sum_{e \in E(u,v)} \frac{\omega(e)}{d(u)}\frac{\gamma_e(v)}{\delta(e)},
\end{align}
where $E(u,v)$ is the set of hyperedges incident to both nodes $u$ and $v$.
Each hyperedge forms a fully connected group of nodes (\cref{fig:hypergraph-reps}c).
Unipartite networks for non-lazy walks have no self-links.
The unipartite representation forms a weighted one-mode projection of the bipartite representation and requires more links with its fully connected groups of nodes.

\paragraph{\textbf{Multilayer.}}
To represent the random walk on a multilayer network, we project the three-stage random-walk process down to a one-step process on state nodes in separate layers. Each hyperedge $e$ with weight $\omega(e)$ forms a layer $A$ with weight $\omega(A)$.
A state node $u^{A}$ represents $u$ in each layer $A \in {E}(u)$ that contains the node.
All state nodes in the same layer form a fully connected set (\cref{fig:hypergraph-reps}d).
The transition rate between state node $u^{A}$ in layer $A$ and state node $v^{B}$ in layer $B$ is
\begin{align}
    P_{uv}^{AB} = \frac{\omega(B)}{d(u)}\frac{\gamma_{B}(v)}{\delta(B)}\text{ for } B \in E(u,v).
\end{align}
With one state node per hyperedge layer that contains the node, the multilayer representation requires the most nodes and links to describe the walk.
Variations include walks that are biased to similar hyperedges~\cite{eriksson_how_2021}.

These network representations and random-walk models have different advantages for modeling multi-body interactions~\cite{eriksson_how_2021,eriksson_flow-based_2022}.
The bipartite network representation enables the most efficient community detection in terms of computation time because it uses the fewest links and often results in shallower solutions.
Multilayer network representations reinforce flows within similar layers, giving the deepest modular structures with the most overlap.
However, this increase in detected modules comes at a high computational cost, as combining fully connected layers with other layers requires more nodes and links than the bipartite network representation.
If the research question does not call for hyperedge assignments or overlapping modules, the unipartite network representation offers a trade-off with intermediate compactness, speed, and the ability to reveal modular regularities.
Among random-walk models, lazy walks typically give more modules in deeper nested structures, while non-lazy walks result in higher modular overlap.

\section{Networks with node attributes} \label{sec:metadata}

Nodes in real-world networks are often labeled with additional information, such as a person's age, an animal's species, an object's shape, or, more generally, as belonging to a certain category.
We can characterize the dynamic process that operates on such an annotated network better by incorporating node-type information in the flow modeling or flow mapping step.

\subsection[Networks with metadata]{\linktonotebook{6.1 Networks with Metadata}{Networks with metadata}}

To combine additional metadata with the network structure, we can modify the flow mapping step by extending the map equation with an additional metadata codebook~\cite{emmons2019pre} or the flow modeling step by choosing which random-walker steps to encode depending on the metadata~\cite{colour-map-equation}. Alternatively,  we can use metadata as a part of an empirical prior for a Bayesian estimate of the random walker’s transition rates~\cite{smiljanic2021comnet} (see \Cref{sec:incomplete}).

For simplicity, we assume unipartite networks and categorical metadata labels and denote node $u$'s label by $\ell_u$.
Further, let $\mathcal{L}$ be the set of possible metadata labels.

\paragraph{\textbf{The content map equation.}}
In addition to communicating the random walker's current node, the content map equation requires encoding the current node's metadata label~\cite{emmons2019pre}.
Given a partition $\mathsf{M}$, the metadata label $\ell$ appears with frequency
\begin{equation}
    r_{\mathsf{m},\ell} = \sum_{u \in \mathsf{m}, \ell_u = \ell} p_u
\end{equation}
within module $\mathsf{m} \in \mathsf{M}$.
The total metadata weight of module $\mathsf{m}$ is 
\begin{equation}
    r_\mathsf{m}^\circlearrowright = \sum_{\ell \in \mathcal{L}} r_{\mathsf{m},\ell} = \sum_{u \in \mathsf{m}} p_u.
\end{equation}
The entropy for encoding the metadata labels for the nodes along the random walker's trajectory in module $\mathsf{m}$ is
\begin{equation}
    H\left(R_\mathsf{m}\right) = - \sum_{\ell \in \mathcal{L}} \frac{r_{\mathsf{m},\ell}}{r_\mathsf{m}^\circlearrowright} \log_2 \frac{r_{\mathsf{m},\ell}}{r_\mathsf{m}^\circlearrowright},
    \label{eq:metadata-entropy}
\end{equation}
where $R_\mathsf{m}$ is the set of metadata frequencies in module $\mathsf{m}$.
Combining \cref{eq:metadata-entropy} with \cref{eq:mapeq} gives the so-called content map equation,
\begin{equation}
    L\left(\mathsf{M}\right) = q_{\curvearrowleft} H\left(Q\right) + \sum_{\mathsf{m} \in \mathsf{M}} p_\mathsf{m}^{\circlearrowright} H\left(P_\mathsf{m}\right) + \eta \sum_{\mathsf{m} \in \mathsf{M}} r_\mathsf{m}^\circlearrowright H\left(R_\mathsf{m}\right), \label{eq:content-map-equation}
\end{equation}
where $\eta$ is a parameter to control the metadata entropy's weight.

\begin{wrapfigure}[27]{r}{.5\linewidth}
    \centering
    \includegraphics[width=\linewidth]{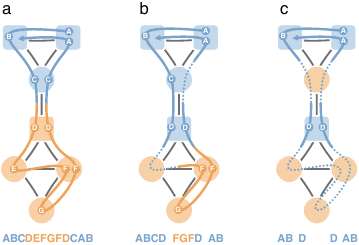}
    \caption{\textbf{Random walks with metadata-dependent encoding probabilities.} 
    In this schematic example with a single, long random walk, we encode the random walker's next step if the target node's metadata are the same as for the previously encoded node.
    When the metadata differ, encoding with a probability between $\frac{1}{3}$ and~$1$ gives the solution in~(a), between~$0.1$ and $\frac{1}{3}$ the solution in~(b), and less than~$0.1$ the solution in~(c).
    Node shapes represent metadata, and node colors optimal partitions.
    The random walks are colored by the currently encoded module and labeled with circles when they encode a transition and dotted lines when they skip nodes.
    The alphabetic codes represent the walks with this metadata-dependent encoding scheme. Figure adapted from Ref.~\cite{colour-map-equation}.}
    \label{fig:metadata-walk}
\end{wrapfigure}

The content map equation produces modules with low metadata entropy. 
It favors splitting structural modules into sub-modules based on their metadata labels and preserving structural boundaries: nodes with the same metadata label remain separated if they belong to different structural modules.

\paragraph{\textbf{Metadata-dependent encoding of random walks.}}
To map nonlocal relationships between metadata and network structure, we use a metadata-dependent encoding of random walks~\cite{colour-map-equation}.
For a random walker starting at node $u$ and stepping to node $v$, we encode the step with probability $\epsilon\left(\ell_u, \ell_v\right)$, which depends on the metadata labels at the start node $u$ and the currently visited node $v$, or the random walker continues without encoding to another node, say $v'$.
We encode the step to $v'$ with probability $\epsilon\left(\ell_u, \ell_{v'}\right)$, or the random walker continues in the same fashion until we eventually encode the step from the start node $u$ to the currently visited node.
After encoding, the random walker restarts at a random node $u'$ chosen proportionally to the node's visit rate.

Varying the coding this way, we derive a metadata-dependent encoding graph from a given network:
The nodes are the same as in the original network, and the links derive from the coding behavior, which can connect disconnected nodes in the original network.
The link weights correspond to the fractions of time that we encode each respective link, such that the weight of link $\left(u, v\right)$ is the number of times that we encode a step to $v$ after restarting at $u$, divided by the total number of encoded steps.
Effectively, this yields a metadata-dependent transition matrix and metadata-dependent visit rates, which we use to detect modules with Infomap.
Unlike the content map equation, modeling metadata-dependent flow enables grouping nodes that belong to different structural modules in the original network.

The encoding probability function determines what specific patterns the metadata-dependent encoding graph captures~(\cref{fig:metadata-walk}).
For example, $\epsilon\left(\ell_u, \ell_v\right) = 1$ forces the walker to encode every step regardless of metadata labels, equivalent to the standard map equation.
Setting $\epsilon\left(\ell_u, \ell_v\right) = \delta_{\ell_u, \ell_v}$, where $\delta$ is the Kronecker delta, encodes only on nodes with the same metadata labels, splitting the network into modules purely based on metadata while ignoring link patterns.
For binary metadata labels, a versatile choice is
%
\begin{equation}
  \epsilon\left(\ell_u, \ell_v\right) = p \delta_{\ell_u, \ell_v} + \frac{p}{c} \left(1 - \delta_{\ell_u, \ell_v}\right),
\end{equation}
%
where $p \in \left[0,1\right]$ and $c \in \left[p, \infty\right]$.
Setting $c > 1$ gives assortative, and setting $p < c < 1$ gives disassortative modules.
For $c = 1$, the expression simplifies to $\epsilon\left(\ell_u, \ell_v\right) = p$ without metadata dependence.
The metadata-dependent encoding probability function can also take real-valued or vectorial metadata labels~\cite{colour-map-equation}.

While generating the metadata-dependent random walk transition graph requires additional computation, the method enables capturing nonlocal interactions, including different metadata types, and modeling different metadata roles.

\subsection[Bipartite networks]{\linktonotebook{6.2 Bipartite Networks}{Bipartite networks}}

The standard map equation can identify modules in bipartite networks, ignoring their two distinct node types.
But by treating these node types, left nodes $V_L$ and right nodes $V_R$, as categorical attributes and making the map equation aware of them, we can exploit the bipartite network’s constrained structure to compress random walks more efficiently.
A bipartite coding scheme captures the alternating pattern of visits between left and right nodes, and the bipartite map equation \cite{PhysRevE.102.052305} improves compression by using separate codebooks for left-to-right and right-to-left transitions,
\begin{align}
    L_B\left(\mathsf{M}\right)
       = q_\curvearrowleft^R H\left(Q^R\right) + \sum_{\mathsf{m} \in \mathsf{M}} p_\mathsf{m}^{R\circlearrowright} H\left(P_\mathsf{m}^R\right)
       + q_\curvearrowleft^L H\left(Q^L\right) + \sum_{\mathsf{m} \in \mathsf{M}} p_\mathsf{m}^{L\circlearrowright} H\left(P_\mathsf{m}^L\right).
    \label{eqn:bipartite-map-equation}
\end{align}
Here, $Q^R$ and $Q^L$ are the left-to-right and right-to-left module entry rates, respectively; $P_\mathsf{m}^R$ and $P_\mathsf{m}^L$ are the sets of right and left node visit rates in module $\mathsf{m}$, including the left-to-right and right-to-left module exit rates, respectively; and $q_\curvearrowleft^R$ and $q_\curvearrowleft^L$ are the rates at which the left-to-right and right-to-left index level codebooks are used, respectively.

To control at which rate the bipartite node types are used, we introduce a node-type forgetting parameter $\alpha$.
Assuming that we misremember node types with probability $\alpha$, the available amount of information about node types is $I = 1-H\left(\alpha, 1-\alpha\right)~\text{bits}$.
Because of this uncertainty about node types, we interpret node visit rates as pairs of left and right flow, leading to mixed visit rates $p_{\alpha,u} = \left(\left(1-\alpha\right)p_u, \alpha p_u\right)$ for left nodes $u \in V_L$, and $p_{\alpha,v} = \left(\alpha p_v, \left(1-\alpha\right) p_v \right)$ for right nodes $v \in V_R$.
The bipartite map equation with varying node-type memory for two-level partitions, $\mathsf{M}$, is
\begin{align}
    L_\alpha\left(\mathsf{M}\right) = q_{\alpha,\curvearrowleft} H\left(Q_\alpha\right) + \sum_{\mathsf{m} \in \mathsf{M}} p_{\alpha,\mathsf{m}}^\circlearrowright H\left(P_{\alpha,\mathsf{m}}\right),
    \label{eqn:bipartite-map-equation-with-varying-node-type-memory}
\end{align}
where $q_{\alpha,\curvearrowleft}$ is the mixed rate at which the index codebook is used, $Q_\alpha$ is the set of mixed module entry rates, and $p_{\alpha,\mathsf{m}}^\circlearrowright$ is the mixed rate at which module $\mathsf{m}$'s codebook is used~\cite{PhysRevE.102.052305}.
As before, we can generalize \cref{eqn:bipartite-map-equation-with-varying-node-type-memory} to multilevel partitions through recursion.
We recover the standard map equation (\cref{eq:mapeq}) for $\alpha=\frac{1}{2}$, and the bipartite map equation (\cref{eqn:bipartite-map-equation}) for $\alpha \in \left\{0,1\right\}$, where setting $\alpha = 1$ flips the node types.
By using node-type information at rates between $I = 0~\text{bits}$ and $I = 1~\text{bit}$, we change the map equation's resolution and detect communities at different scales.

By varying the amount of available node-type information, $I$, the bipartite map equation reveals the organizational structure of bipartite networks at different resolutions \cite{PhysRevE.102.052305}.
\section[Networks with incomplete data]{\linktonotebook{7 Networks with incomplete data}{Networks with incomplete data}} \label{sec:incomplete}

In the previous sections, we discussed how to take advantage of the available data and identify flow models that best describe modular regularities. 
Once we have chosen a flow model, the standard map equation estimates the stationary flow distribution based on what it sees in the data. 
However, data can be noisy with incomplete link measurements, leading to inaccurate estimates of the actual stationary flow distribution in the system. 
Consequently, when missing observations distort the balance between module and index-level codebooks, the map equation can overfit to the noise and highlight spurious communities~\cite{ghasemian2019overfitting, smiljanic2020pre, smiljanic2021comnet}. 

\begin{figure}[htp!]
    \centering
    \includegraphics[width=0.7\linewidth]{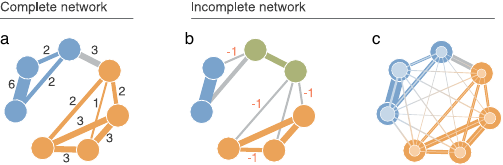}
    \caption{\textbf{Community detection in an undirected weighted network with complete and missing link observations.}
    (a) The map equation identifies two modules in a complete network with accurate network flows.
    (b) Missing link observations introduce inaccuracies leading to overfitting.
    (c) The map equation with regularized network flows recovers the complete network’s community structure. Colors represent community assignments. The width of the lighter lines represents prior link weights, and the size of the lighter node centers indicates prior transition probability. Figure adapted from Ref.~\cite{smiljanic2021comnet}.}
    \label{fig:regularization}
\end{figure}

To address this challenge, we can extend the map equation to account for incomplete data by applying a Bayesian method to regularize the random walker's transition rates (\cref{fig:regularization}) \cite{smiljanic2021comnet}.
This method allows us to account for potentially missing connections in the observed network. It combines the observed connections with prior expectations, effectively reducing the effects of sampling noise.

\subsection{Bayesian estimate of transition rates}
We assume that a random walker steps from node $u$ to nodes $v$ proportional to $\tilde{p}_u=(\tilde{p}_{u v_1}, \dots, \tilde{p}_{u v_N})$.
The link weights $(w_{u v_1}, \dots, w_{u v_N})$ represent a sample of the hidden distribution $\tilde{p}_u$. 
We assume non-negative integer weights for simplicity, but the method also works for non-negative real weights.
If we interpret the transition probability $p_{uv}$ given by \cref{eq:transitionrate} as a maximum likelihood estimate of $\tilde{p}_{uv}$, $p_{uv}$ can deviate significantly from $\tilde{p}_{uv}$ in the case of noisy data.
To reduce the effects of statistical fluctuations, we estimate the average posterior transition rates
\begin{align}
 \hat{p}_{uv}\left(w\right) = \int \tilde{p}_{uv} P(\tilde{p}_u | w_{u v_1}, \dots, w_{u v_N}) d\tilde{p}_u, \label{eq:bayes_estimate}
\end{align}
where $P(\tilde{p}_{u} | w_{u v_1}, \dots, w_{u v_N})$ is a posterior over the unknown distribution $\tilde{p}_{u}$ given by Bayes' rule.
As prior distribution $P(\tilde{p}_{u})$, we choose the Dirichlet distribution, which is the conjugate prior of the multinomial distribution and enables analytical calculations:
\begin{align}
 P(\tilde{p}_u | \gamma_u) = \frac{\Gamma(\gamma_{u v_1} + \dots + \gamma_{u v_N})}{\Gamma(\gamma_{u v_1}) \ldots \Gamma(\gamma_{u v_N})} \prod_{v \in V} \tilde{p}_{uv}^{\gamma_{uv} - 1}.
\end{align}
$\Gamma(x)$ is the gamma function, and $\gamma_{u v_1}, \ldots, \gamma_{u v_N}$ are parameters of the distribution that reflect our prior assumption about links before we observe the network data.
After integrating \cref{eq:bayes_estimate}, we obtain
\begin{align}
 \hat{p}_{uv} = (1 - \alpha_u)\frac{w_{uv}}{\sum_{v}w_{uv}} + \alpha_u \frac{\gamma_{uv}}{\sum_{v}\gamma_{uv}}, \label{eq:bayes_transition_rate}
\end{align}
where
\begin{equation}
  \alpha_u = \frac{\sum_{v \in V} \gamma_{uv}}{\sum_{v \in V} w_{uv} + \gamma_{uv}}.
\end{equation}
%
%
The Bayes estimate of the transition rates in \cref{eq:bayes_transition_rate} represents a weighted average of the empirical transition probabilities and the prior expectations. As the number of observed out-links of node $u$ increases, the estimate approaches the empirical probabilities, whereas sparse observations cause the prior parameters to dominate the estimate. Thus, the quality of this Bayesian estimate depends both on the number of link observations and on the choice of the prior parameters $\gamma_{uv}$.

\subsection{Prior parameter}
When selecting prior parameters $\gamma$, we aim to ensure that the map equation identifies meaningful communities when enough data are provided, while avoiding overfitting in undersampled networks. 
In the limit of very sparse data, a desirable outcome is a single community encompassing all nodes, reflecting that the available observations do not support finer partitions. 

We treat link probabilities and link weights as decoupled and write
\begin{align}
  \gamma_{uv} = \lambda_{uv} c_{uv},
\end{align}
where $\lambda_{uv}$ is a connectivity parameter that denotes the prior probability that nodes $u$ and $v$ are connected, and the weight parameter $c_{uv}$ determines prior link weights.

\paragraph{\textbf{Prior link weights.}}
We use the so-called continuous configuration model~\cite{palowitch2018jmlr} to obtain an empirical estimate for the prior link weights,
\begin{equation}
    c_{uv} = \frac{\sum_{n \in V} k_n^{\text{in}} + k_n^{\text{out}}}{\sum_{n \in V} s_n^{\text{in}} + s_n^{\text{out}}} \frac{s_u^{\text{out}} s_v^{\text{in}}}{k_u^{\text{out}} k_v^{\text{in}}},
    \label{link_weight}
\end{equation}
where $k_u^{\text{in}}$ and $k_u^{\text{out}}$ denote observed in- and out-degrees, and $s_u^{\text{in}} = \sum_v w_{vu}$ and $s_u^{\text{out}} = \sum_v w_{uv}$ denote in- and out-strengths for node $u$.
The continuous configuration model preserves the expected weights of in- and out-links.
For unweighted networks, where $s_u=k_u$, it assigns weights $c_{uv}=1$ to all links.

\paragraph{\textbf{General connectivity parameter.}}
To reduce the bias induced by incomplete observations, we assume an uninformative connectivity corresponding to a network without modular structure, where nodes $u$ and $v$ are connected with uniform probability.
If no information about node attributes is provided, we use
\begin{equation}
  \lambda_{uv}= \lambda = \frac{\ln N}{N}.
\end{equation}
This value for $\lambda$ is the theoretical threshold at which Erd{\H{o}}s-R{\'e}nyi random networks become almost surely connected.
When $\lambda < \frac{\ln N}{N}$, the prior network has disconnected components and can fail to reduce overfitting, while $\lambda > \frac{\ln N}{N}$ can be too strong and wash out modular regularities~\cite{smiljanic2020pre, smiljanic2021comnet}.

\begin{figure*}[bp!]
    \centering
    \includegraphics[width=\linewidth]{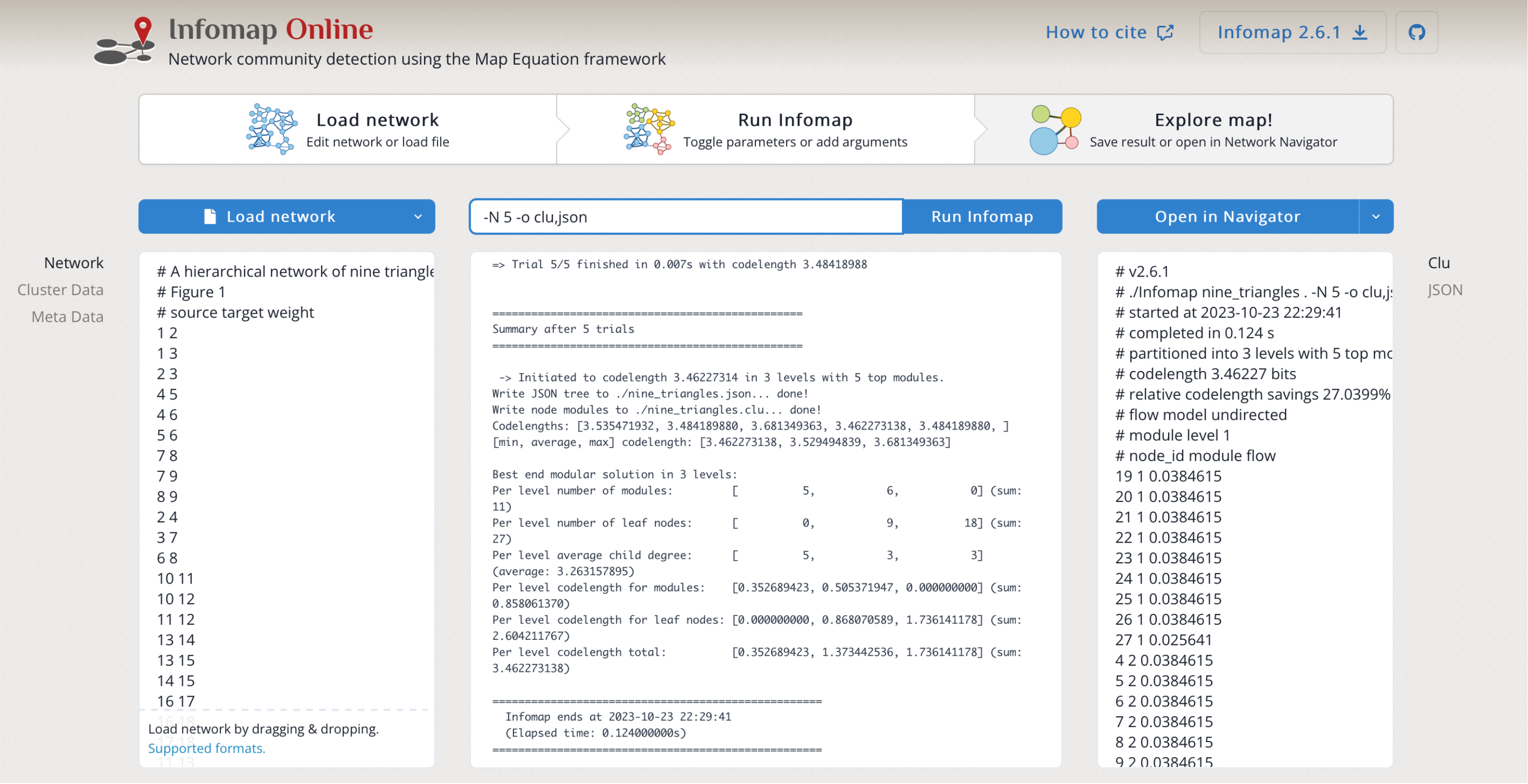}
    \caption{\textbf{Screenshot of Infomap Online}.
    Infomap Online is a client-side web application that enables users to run Infomap in the web browser.
    We achieve this by compiling Infomap from \CC\ to JavaScript and provide a graphical user interface for all input and output parameters.
    Infomap Online is available at \url{https://www.mapequation.org/infomap/}.}
    \label{fig:infomap-screenshot}
\end{figure*}

\paragraph{\textbf{Bipartite connectivity parameter.}}
If the given network has a bipartite structure with $N_L$ left nodes and $N_R$ right nodes, we adjust the connectivity parameter accordingly: we use the smallest $\lambda^\text{bi}$ at which a random bipartite network is almost surely connected~\cite{saltykov1995discretemath}
\begin{equation}
  \lambda^\text{bi} = \frac{\ln \left(N_L + N_R\right)}{\min \left(N_L, N_R\right)}.
\end{equation}
We also adapt the definition of prior link weights to reflect that links in bipartite networks must connect different-type nodes,
\begin{equation}
  \gamma_{uv}^\text{bi} = \left(1 - \delta_{\tau_u \tau_v}\right) \lambda^\text{bi} c_{uv},
\end{equation}
where $\tau_u$ and $\tau_v$ are the types of nodes $u$ and $v$, that is, left or right.

\paragraph{\textbf{Metadata connectivity parameter.}}
Nodes annotated with metadata allow further fine-tuning the connectivity parameter.
We expect the $N_\ell$ nodes with the same metadata label $\ell$ to be connected with higher probability.
With metadata labels $\ell_u$ and $\ell_v$ for nodes $u$ and $v$, we adjust the prior connectivity to
\begin{equation}
  \lambda_{uv}^{\text{meta}} = \lambda + \delta_{\ell_u \ell_v} \lambda_{\ell_u},
\end{equation}
where $\delta$ is the Kronecker delta, and $\lambda_{\ell_u} = \frac{\ln N_{\ell_u}}{N_{\ell_u}}$.

\section{Software and visualization} \label{sec:visualization}
Infomap is the search algorithm and a key component of the map equation framework.
But to understand the organization of large complex networks, we require both efficient algorithms and powerful visualizations.
We have developed several visualization tools with different purposes and made them available as web applications for anyone to use.
The first step of community detection with the map equation framework is to install or get access to Infomap.

\subsection{Infomap}
Infomap is the name of the search algorithm that optimizes the map equation.
Its core is written in \CC\ and requires a compiler with \CC-14 support.

\paragraph{\href{https://pypi.org/project/infomap/}{\textbf{Python package~\faExternalLink}}}
To support the many researchers that use Python for their community detection pipelines, we made Infomap available as a Python package, also called \texttt{infomap}~\cite{mapequation2020infomappython}.
This package is a small wrapper around the core \CC\ version of Infomap with additional helper functions and integrates with other popular packages, such as \texttt{networkx} and \texttt{pandas}.
A basic version of the package is also included in the popular community detection library CDLib~\cite{rossetti2019cdlib}.

\paragraph{\href{https://www.mapequation.org/infomap/}{\textbf{Infomap Online~\faExternalLink}}}
\begin{wrapfigure}{r}{.42\linewidth}
    \centering
    \includegraphics[width=\linewidth]{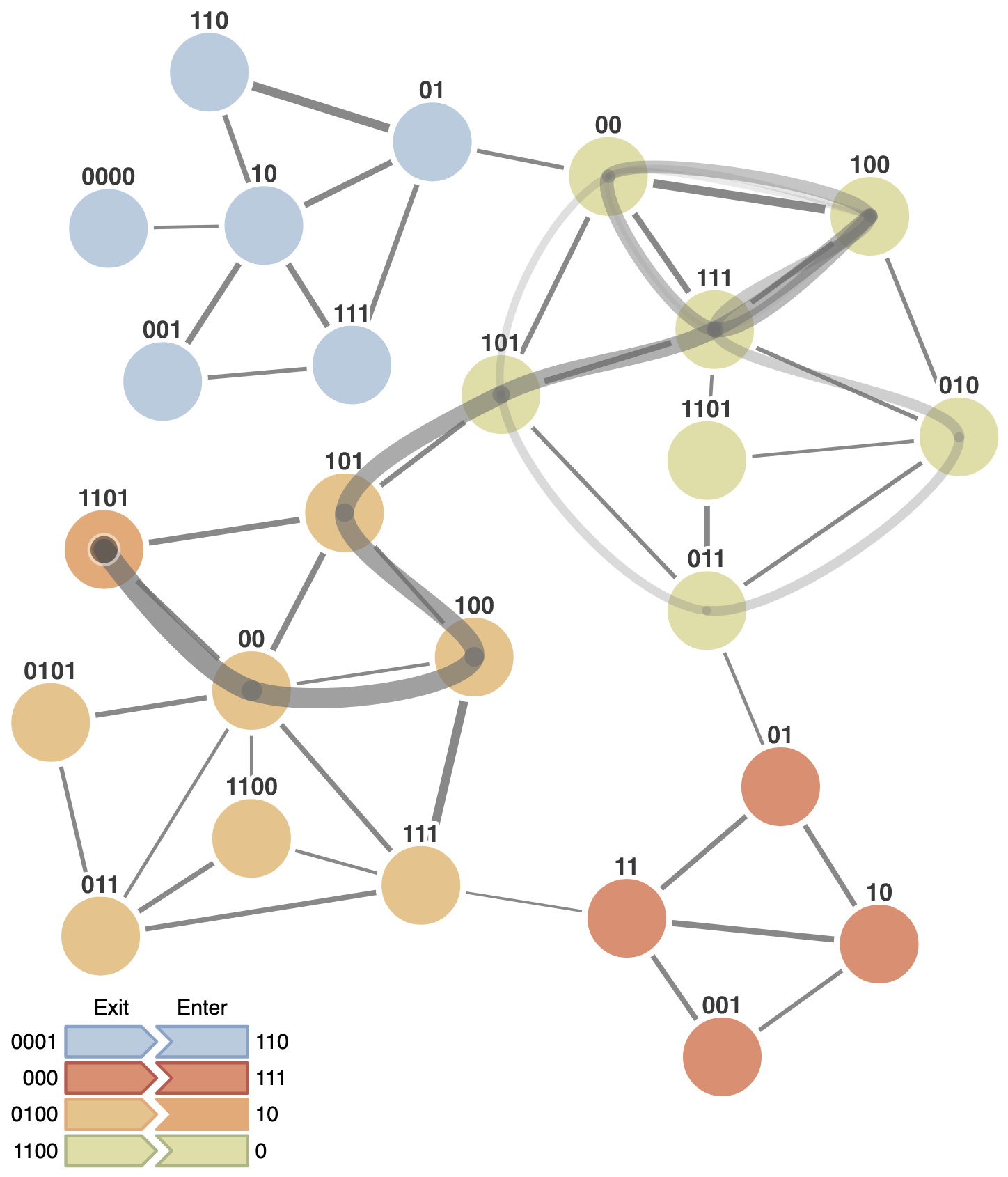}
    \caption{\textbf{Modular description of a random walk on a network.}
    Each module has its own codebook comprised of node and exit codewords.
    A global index codebook is comprised of enter codewords for all modules.
    Each codeword is selected using the Huffman code algorithm on the possible transitions in each codebook.}
    \label{fig:map-demo}
\end{wrapfigure}
While the most computationally efficient community detection pipeline is to run the stand-alone \CC\ or Python versions of Infomap, we have embedded a JavaScript version of Infomap into a web application we call Infomap Online~(\cref{fig:infomap-screenshot})~\cite{mapequation2022infomaponline}.
This embedded Infomap version supports the same network inputs as \CC\ Infomap.
Infomap Online also serves as interactive documentation for Infomap with examples and simple network visualizations.

\subsection[Map equation demo]{\href{https://www.mapequation.org/demo/}{Map equation demo~\faExternalLink}}

While the map equation does not depend on explicitly deriving codewords or simulating random walks, visualizing the modular description of a random walk can highlight the mechanics of the map equation.
To do this, we have developed a web-based demo application~\cite{mapequation2022demo}.
In this application, we show how modular Huffman codes~\cite{huffman_method_1952} enable re-using codewords in different modules by encoding module transitions with exit and enter codewords.
This enables a shorter expected codelength for partitions where the random walker spends a relatively long time inside modules compared to exiting and entering other modules.
In \cref{fig:map-demo}, we show a realization of a random walk starting in the green -- top right -- module and transitioning to the orange -- bottom left -- module.
Every module transition is encoded using a module-specific exit codeword followed by an enter codeword from the index codebook.

\subsection[Infomap Network Navigator]{\href{https://www.mapequation.org/navigator/}{Infomap Network Navigator~\faExternalLink}}

\begin{figure}[htp!]
    \centering
    \includegraphics[width=\linewidth]{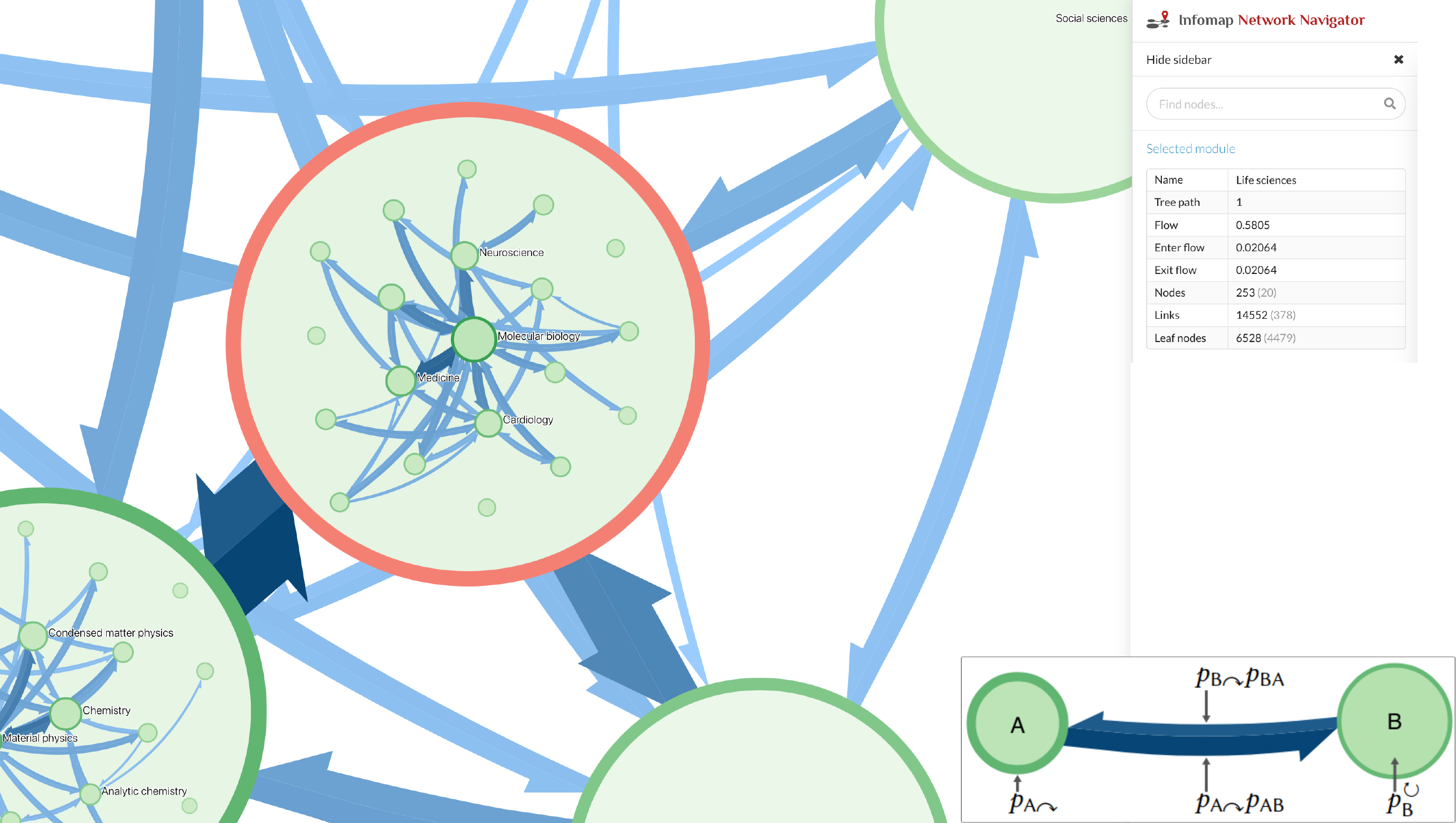}  
    \caption{\textbf{Infomap Network Navigator.}
    The Network Navigator is a web application that generates an interactive zoomable map for networks clustered with Infomap.
    Main panel: The citation flow patterns in science are organized in a multilevel partition.    
    Zooming in reveals sub-modules within the life sciences, such as molecular biology and medicine.
    Inset: Schematic representation of two connected modules. Modules are represented by circles with areas proportional to the contained flow $p_{\mathsf{m}\circlearrowright}$ and border thickness proportional to the exiting flow $p_{\mathsf{m}\curvearrowright}$.
    Links with thickness and lightness proportional to -- and length inversely proportional to -- the aggregated flow volume between modules $p_{m\curvearrowright} p_{mn}$.    
    The Infomap Network Navigator is available at \url{https://www.mapequation.org/navigator/}.
    }
    \label{fig:navigator}
\end{figure}

Understanding real-world systems relies on effectively mapping complex networks.
To comprehend and navigate the intricate multilevel communities within these networks, it is crucial to employ powerful visualizations that can transform textual representations into visually intuitive maps.
To address this need, we have developed an interactive web application called Infomap Network Navigator~\cite{mapequation2018navigator,eriksson2018interactive}.
This tool allows users to visualize and explore multilevel maps of both conventional and higher-order networks in a manner similar to using mapping software such as Google Maps.

We visualize multilevel communities using top-level modules represented by circles.
The circle areas indicate the contained flow volume, while the border thickness represents the exiting flow volume~(inset in \cref{fig:navigator}).
Modules are connected by aggregated links, with their thickness and color lightness reflecting the flow between them.
Additionally, we adjust the link lengths inversely proportional to the inter-module flow, emphasizing varying connection strengths.
\Cref{fig:navigator} illustrates using the Infomap Network Navigator to highlight multilevel citation flow patterns in science.
At the top level, the application visualizes research fields like the life sciences, physical sciences, and social sciences.
By zooming in using the mouse wheel or trackpad, more detailed information becomes visible.
For example, within the life sciences, the application reveals further divisions into research areas such as molecular biology and medicine.

\subsection[Alluvial Diagram Generator]{\href{https://www.mapequation.org/alluvial/}{Alluvial Diagram Generator~\faExternalLink}}

\begin{figure*}[htp!]
    \centering
    \includegraphics[width=0.82\linewidth]{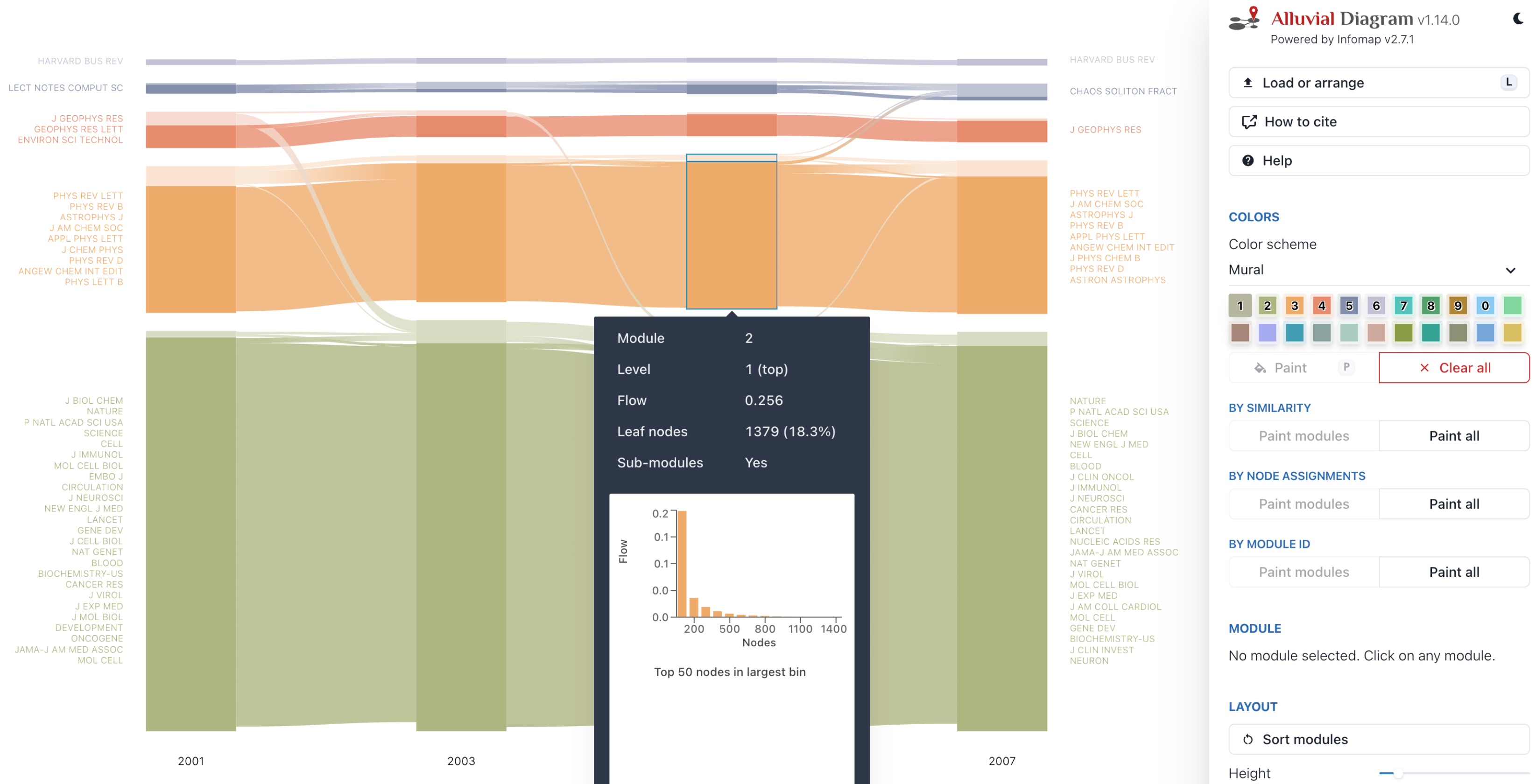}
    \caption{\textbf{Screenshot of the Alluvial Diagram Generator}. 
    Alluvial diagrams visualize how the modular organization of networks changes. 
    Each vertical stack represents the network's organization, here for journal citation data from different time points.
    The Alluvial Diagram Generator is available at \url{https://www.mapequation.org/alluvial/}.}
    \label{fig:alluvial-screenshot}
\end{figure*}

Complex systems, such as social or biological systems, are dynamic.
This results in changing interaction patterns over time.
Understanding this change is essential to gain insights into the organization and evolution of complex systems.

\begin{wrapfigure}[26]{r}{.45\linewidth}
    \includegraphics[width=\linewidth]{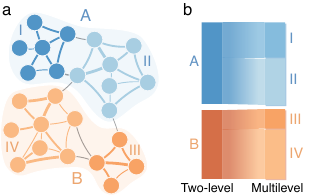}
    \caption{\textbf{Schematic alluvial diagram of multilevel network structure.}
    (a) A weighted network with a multilevel modular structure.
    The two blue and orange top-level modules (\textsf{A--B}) are further divided into two sub-modules each (I--IV), indicated by color lightness.
    (b) An alluvial diagram representation of the top-level and multilevel modules in (a) using the same colors.
    Columns of blocks represent modules with heights proportional to the contained flow volume.
    The leftmost column is an ordinary two-level alluvial diagram representation.
    The multilevel representation to the right shows multiple levels, with the background showing the top-level organization.
    Streamfields connect modules in the left and right columns that share nodes.}
    \label{fig:schematic-alluvial}
\end{wrapfigure}

Various summary statistics can quantify these structural changes, such as variation of information or normalized mutual information, which compute a distance or similarity between two partitions~\cite{lancichinetti_detecting_2009, good_performance_2010, esquivel2012comparing, newman_improved_2020}.
While these metrics have their use cases, they do not capture essential information about how the networks change.

Powerful visualization tools are necessary to effectively visualize and understand how and where these changes occur.
Alluvial diagrams visually represent the changing organization of networks by embedding each network's modules as vertically stacked blocks (\cref{fig:alluvial-screenshot}) \cite{rosvall_mapping_2010, holmgren2023mapping}.
The height of these blocks is proportional to the flow volume of the nodes they contain.
By positioning different networks adjacent to one another and connecting modules with shared nodes using streamfields -- the middle part of \cref{fig:schematic-alluvial}b -- the diagrams highlight the changing organization.
To incorporate multilevel solutions in alluvial diagrams, we display hierarchically nested sub-modules above their corresponding super-modules.
We illustrate this in the right stack in \cref{fig:schematic-alluvial}b.

We developed an alluvial diagram generator as a web application~\cite{mapequation2022alluvial} that operates within the user's web browser, ensuring that data never leave the local machine.
This client-side architecture is crucial for researchers working with sensitive data.
\section{Applications} \label{sec:applications}

Applications in network science range from fundamental tasks such as assessing node centrality and similarity to more advanced approaches, including bioregionalization and model selection for correlational data.
These problems are often tackled with heuristic methods that provide practical but ad hoc solutions.
By capitalizing on the map equation's coding principles, we can instead ground them in a principled, information-theoretic framework.
With Infomap, this foundation translates into efficient and robust methods that support reliable applications across fields from biodiversity research to systems biology.

\subsection[Map equation centrality]{\linktonotebook{9.1 Map Equation Centrality}{Map equation centrality}}
Node centrality measures characterize how important or influential each node in a network is.
Traditional measures take either microscopic  or macroscopic perspectives: node degree centrality counts local connections, while PageRank tracks global flows, for example.
Yet these measures overlook the mesoscopic community structure.
Degree centrality cannot distinguish between a node embedded deep within a community and a bridge node if they have the same number of connections, just as PageRank cannot if they have identical visit rates.

Drawing on the map equation's coding principles, map equation centrality offers an information-theoretic approach to quantifying community-aware node importance.
Similar to how certain auctions calculate the winner's price as the total harm it causes other bidders \cite{vickrey1961counterspeculation,leonard1983elicitation}, map equation centrality quantifies a node's importance as the collective benefit to other nodes if that node were removed from the coding scheme.
Specifically, how many bits shorter the codewords for remaining nodes would become without having to encode visits to the target node \cite{map-equation-centrality}.

Given a partition $\mathsf{M}$, map equation centrality for node $u$ in module $\mathsf{m}_u$ can be expressed as
\begin{equation}
    \lambda\left(\mathsf{M}, u\right) = -\left(p_{\mathsf{m_u}}^\circlearrowright - p_u\right) \log_2 \frac{p_{\mathsf{m}_u}^\circlearrowright - p_u}{p_{\mathsf{m}_u}^\circlearrowright}.
\end{equation}
Because the map equation uses modular codebooks, removing a node only affects codewords within its own module, making the centrality calculation locally efficient.

In networks with clear community structure, map equation centrality reveals patterns invisible to traditional measures.
For example, bridge nodes connecting communities may have modest degree or PageRank, yet their removal can drastically decrease coding costs for remaining nodes in the module, revealing their centrality from a community-aware perspective.

\subsection[Map equation similarity]{\linktonotebook{9.2 Map Equation Similarity}{Map equation similarity}}
Node similarity measures quantify how similar two nodes are, enabling representation learning and link prediction in networks.
Recent approaches learn features from local subgraphs, overlapping neighborhoods, or shortest paths to embed nodes in low-dimensional Euclidean spaces~\cite{grover_node2vec_2016}.
But symmetric distance measures cannot distinguish directed relationships where similarity from $u$ to $v$ differs from $v$ to $u$, and low-dimensional representations struggle to capture complex, non-metric relationships in networks.

Leveraging the map equation's coding principles, map equation similarity -- \emph{mapsim} for short -- sidesteps these geometric constraints \cite{map-equation-similarity}.
Instead of embedding nodes in metric space, mapsim grounds similarities in compression efficiency within the network's modular structure.
While networks constrain random walks to existing links, the map equation's coding scheme can describe transitions between any node pair.
Once we know a network's community structure and its corresponding codewords, we can calculate the description length for any potential transition, whether the link exists or not.

To calculate how similar node $u$ is to node $v$ given partition $\mathsf{M}$, $\operatorname{mapsim}\left(u, v, \mathsf{M}\right)$, we find the smallest module $\mathsf{m} \in \mathsf{M}$ that contains both $u$ and $v$.
The similarity equals the product of the rate $r_{u,\mathsf{m}}$ at which a random walker at $u$ transitions up to the root level of $\mathsf{m}$ and the rate $r_{\mathsf{m},v}$ at which a random walker at the index level of $\mathsf{m}$ transitions to and visits node $v$,
$\operatorname{mapsim}\left(u, v, \mathsf{M}\right) = r_{u,\mathsf{m}} r_{\mathsf{m},v}$.
We interpret $\operatorname{mapsim}\left(u, v, \mathsf{M}\right)$ as the distance from $u$ to $v$, in bits, by setting $d_{uv} = \log_2 \operatorname{mapsim}\left(u, v, \mathsf{M}\right)$.

This information-theoretic, community-based approach proves effective: across 47 networks, mapsim achieved average performance more than 7 percent higher than its closest competitor \cite{map-equation-similarity}.

\subsection[Infomap Bioregions]{\linktonotebook{9.3 Infomap Bioregions}{Infomap Bioregions}}

\begin{figure}[htp!]
    \centering
    \includegraphics[width=0.7\linewidth]{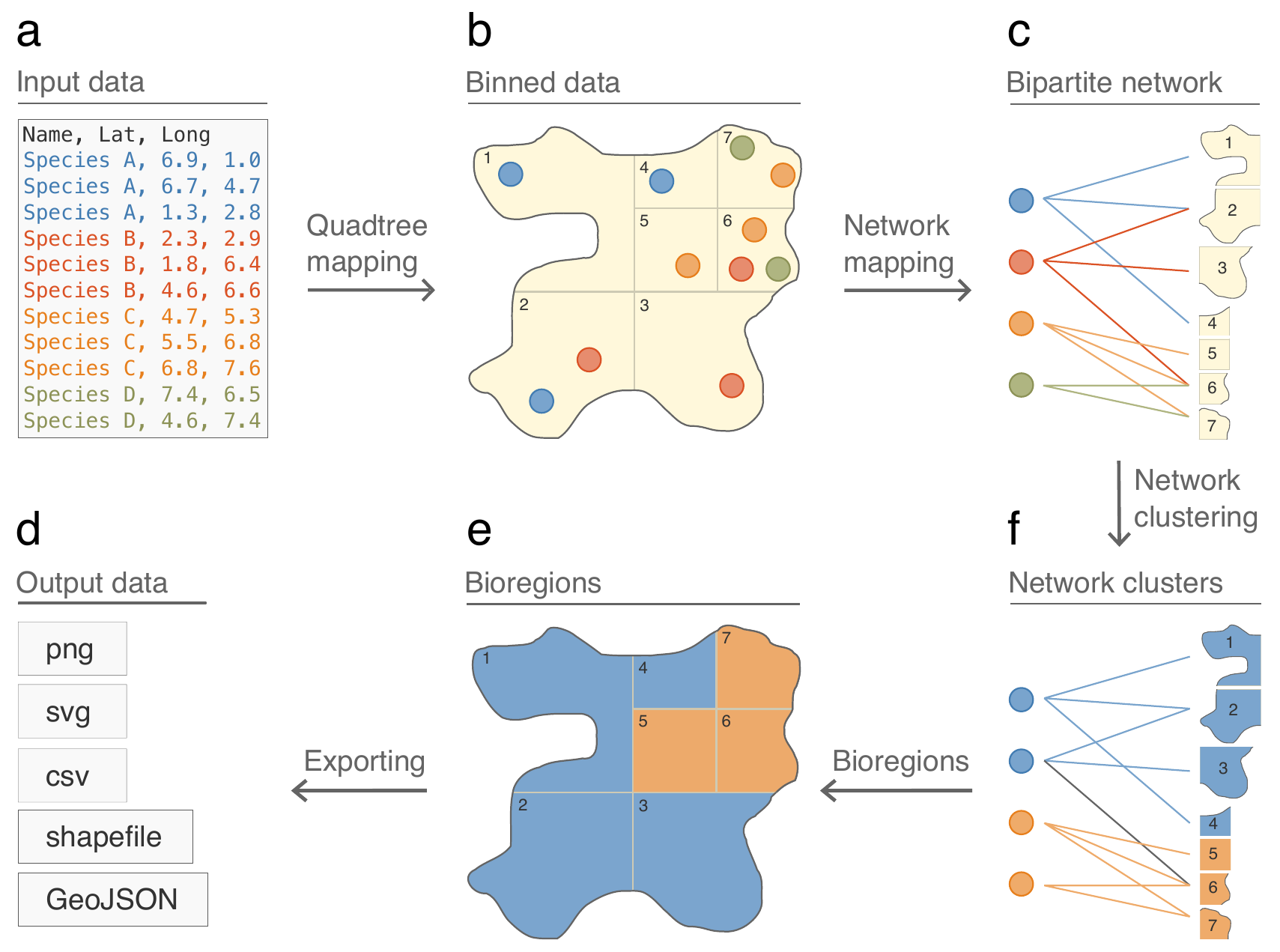}
    \caption{\textbf{Infomap Bioregions}.
    Schematic illustration of bioregional mapping from species distribution data to bioregions.
    Network-based methods identify bioregions from species distribution data (a) by binning species observations into spatial grid cells (b).
    Grid cells and species form nodes in a bipartite network where a links indicate species occurrence in grid cells (c).
    Infomap detects communities (d).
    In bipartite networks, communities contain grid cells and species; the grid cells assigned to the same community form a bioregion (e).
    In the interactive Infomap Bioregions tool, available at \url{https://www.mapequation.org/bioregions/},
    bioregional maps can be exported to various formats (f).}
    \label{fig:bioregions-schematic}
\end{figure}

\begin{figure*}[htp!]
    \centering
    \includegraphics[width=0.9\linewidth]{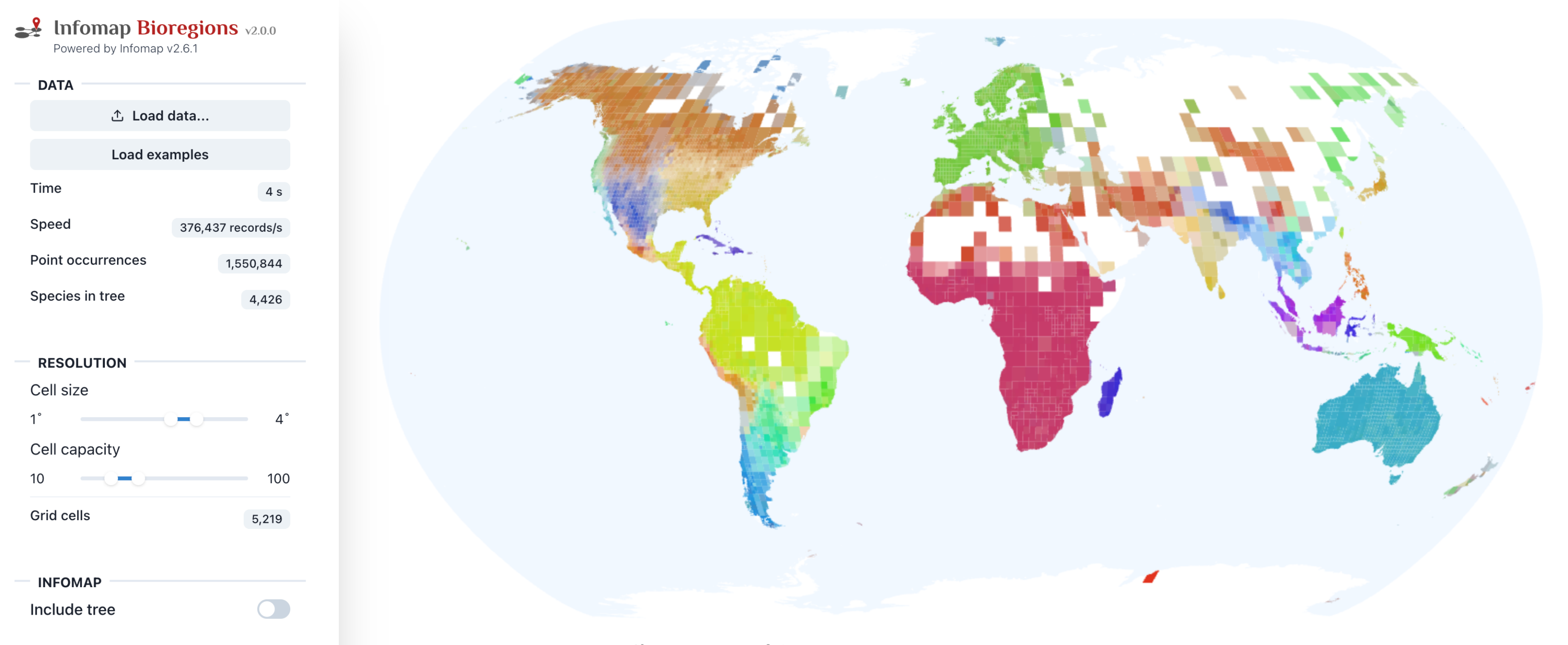}
    \caption{\textbf{Screenshot of Infomap Bioregions}.
   The web application enables interactive exploration of species distribution data and bioregional patterns.
    The map shows mammal occurrence distributions where white areas lack sufficient data for statistics.
    Infomap Bioregions is available at \url{https://www.mapequation.org/bioregions}.}
    \label{fig:bioregions-screenshot}
\end{figure*}

\noindent In biodiversity research, data-derived biogeographical regions, or simply bioregions, reveal how species are grouped at large spatial scales.
They serve as essential units for understanding historical biogeography, ecology, and evolution, and for identifying the areas most critical for conservation.
Traditional approaches rely on subjective clustering of similarity measures, often missing the complex co-occurrence patterns that define ecological communities. 

Infomap Bioregions grounds bioregional delimitation in modular compression of network flows \cite{edler_infomap_2017}.
The method first bins all species into spatial grid cells. These species and grid cells form a bipartite network, where links connect species to grid cells where they occur.
Infomap then identifies bioregions by detecting network modules where species assemblages remain distinct, revealing natural boundaries in biological diversity (\cref{fig:bioregions-schematic}).
The interactive web application allows researchers to upload occurrence data, tune parameters, and export results for further analysis (\cref{fig:bioregions-screenshot})

This principled approach has proven effective for conservation biologists. They have used Infomap Bioregions to delineate bioregions in one of our planet's most biodiverse hotspots, the tropical Andes, to understand the spatial and temporal evolution of the biota \cite{hazzi2018biogeographic}.
In another study, researchers used this approach to analyze about 25,000 vascular plant species in tropical Africa, determining whether different forms of plant growth display similar diversity patterns \cite{droissart2018beyond}.

\subsection[Model selection for correlational data]{\linktonotebook{9.4 Model Selection with Correlational Data}{Model selection with correlational data}}

\begin{figure*}[htp!]
    \centering
    \includegraphics[width=\linewidth]{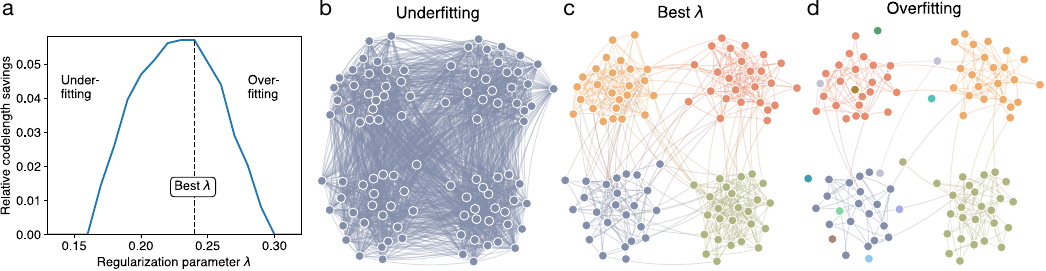}
    \caption{\textbf{Model selection with the map equation.}
    We can use the map equation to find the best network model of correlational data or other similarity data. The relative codelength savings show how much the model's description length is compressed as we vary regularization strength, comparing the best identified partition to a one-module solution. The codelength savings peak at the optimal model (a), which best balances under- and overfitting modular structure to the data, as illustrated by  weak (b), optimal (c), and strong (d) regularization.}
    \label{fig:model-selection}
\end{figure*}

\noindent Correlational data represented as networks underpin analyses across the sciences.
Identifying modules in these networks can reveal genes with shared functions or species that prefer similar conditions.
But they are typically dense and noisy, making it difficult to distinguish meaningful structure from statistical artifacts.
To extract useful information from the correlation networks, researchers must regularize them, but common approaches based on heuristics or likelihood maximization risk either erasing structure by underfitting or generating spurious patterns by overfitting.

The map equation offers a more principled approach (\cref{fig:model-selection}).
By aligning model selection directly with the goal of identifying modular structure, it replaces arbitrary thresholding with a direct measure of clustering performance.  
Specifically, the map equation's maximum codelength savings determine the regularization strength.
This balance between complexity and fit helps identify only those modules that are robust to noise and supported by data \cite{Neuman2022, Neuman2023, Neuman2025}.
This principle extends to Gaussian graphical models. A module-based graphical lasso infers sparse precision matrices from correlational data with improved performance under noise \cite{Neuman2023}.

Applied to gene co-expression data, the method uncovers modular structure that remain hidden with standard techniques such as WGCNA or the graphical lasso \cite{Neuman2022,Neuman2023,Neuman2025}.
\section{Conclusion} \label{sec:conclusion}
Simplifying the dynamic processes on networks to uncover how complex systems work remains a challenging problem with many applications.
To help researchers take full advantage of the map equation framework and its community-detection algorithm Infomap, we explain the information-theoretical principles of the map equation and summarize its many generalizations to various network representations, flow models, and modular descriptions.
As network science advances, flow-based community detection methods will remain essential for unraveling the complexities of networked systems and their inner workings.
We hope our review can inspire further generalizations of the map equation to simplify and highlight important structures in innovative network models.

\subsection{Data and code availability}

Software is available at \url{https://mapequation.org}.
Infomap is available at \url{https://mapequation.org/infomap}
and its source code at \url{https://github.com/mapequation/infomap}
Data and notebooks are available at \url{https://github.com/mapequation/infomap-tutorial-notebooks}.

\subsection{Author contributions}
All authors wrote and edited the manuscript.

\subsection{Competing interests}
The authors declare that they have no competing interests.

\begin{acks}
J.S., D.E., and M.R. were supported by the Swedish Research Council, Grant No.\ 2016-00796.
C.B.\ was supported by the Wallenberg AI, Auto\-no\-mous Systems and Software Program (\href{https://wasp-sweden.org}{WASP}) funded by the Knut and Alice Wallenberg Foundation, the German Federal Ministry of Education and Research, Grant No.\ 100582863 (TissueNet), and the Swiss National Science Foundation, Grant No.\ 176938.
A.H.\ and M.N.\ were supported by the Swedish Foundation for Strategic Research, Grant No.\ SB16-0089.
\end{acks}

\bibliographystyle{ACM-Reference-Format}
\bibliography{references}

\end{document}